# Modelling Socio-ecological Systems: Implementation of an Advanced Fuzzy Cognitive Map Framework for Policy development for addressing complex real-life challenges

Mamoon Obiedat[1] and Sandhya Samarasinghe[2]

**Abstract** This study implements a novel Fuzzy Cognitive Map (FCM) framework for addressing large complex socio-ecological problems. These problems are characterized as qualitative, dominated by uncertainty, human involvement with different and vague perceptions/expectations, and complex systems dynamics due to many feedback relations. The FCM framework, proposed previously by us, and with enhancements made in this study, provides a participatory soft computing approach to develop consensus solutions. We demonstrate its implementation in a case study- a national-scale acute water scarcity crisis. The model has eight steps starting from collecting data from stakeholders in the form of FCMs (bi-directional graphs) represented by nodes and imprecise connections. All subsequent steps operate within a new fuzzy 2-tuple framework that overcomes previous FCM limitations through advanced processing methods, where large FCMs are fuzzified and analyzed, condensed, and aggregated using graph-theoretic measures. FCMs are simulated as Auto-Associative Neural Networks (AANN) to assess policy solutions to address the problem. In this study, very large cognitive maps were developed through interviews capturing perceptions of five different stakeholder groups taking into consideration the causes and consequences as well as emerging challenges of the acute water scarcity problem in Jordan. The complex FCMs containing 186 variables comprehensively covered all aspects of water scarcity. These FCMs were condensed into smaller maps in two levels. They were also combined into five stakeholder group FCMs and one whole system FCM (total 123 FCMs). AANN simulations of policy scenarios were conducted on the whole system FCM, first at the most condensed level and then moved top-down through the next two levels of granularity to explore potential solutions. These were ranked by a novel fuzzy *Appropriateness* criterion to select the most feasible solutions.  This systematic approach provided a number of high level and effective strategies to mitigate the water crisis.

[1] Department of Information Technology, The Hashemite University, Jordan.
   mamoon@hu.edu.jo, ORCID: 0000-0003-3151-9043

[2] Complex Systems, Big Data and Informatics Initiative (CSBII), Lincoln University, Christchurch, New Zealand.
    Sandhya.Samarasinghe@lincoln.ac.nz, ORCID: 0000-0003-2943-4331

**Keywords** Fuzzy cognitive maps, fuzzy 2-tuple, Complexity, Socio-ecological systems, Water scarcity, Policy simulations

## 1. Introduction

Social-ecological or environmental engineering systems involve human-environment interactions that are by nature uncertain, complex, dynamic, qualitative and participatory. Comprehending and addressing problems in these systems (*e.g.*, global water problems) requires a comprehensive insight including both human and ecological dimensions. In such systems, humans (the social system) interact greatly with the environment (the ecosystem) in complex ways, as shown in Fig. 1. To represent these dimensions of reality that typically exist in an environment of conflict in a systems model requires, in addition to documented knowledge, the human knowledge existing at all levels of skill and perception [1]. Yet, collecting and modelling human perceptions is a challenge because human reasoning is vague and variable producing ambiguous evaluations and uncertain data/information [2]. However, human participation in problem-solving in these domains is truly required and hence, soft computing solutions are needed to tackle these challenging issues.

A domain where a participatory approach to problem solving has been acutely felt is water resources engineering. According to current trends and prospects for clean water, global water demand will exceed existing water resources within a few decades, threatening the world's water shortage [3]. A case in point is Jordan which is one of the countries most affected by water scarcity in the world [4, 5]. Due to the severe increase in water demand and the decrease in renewable water supply, the annual per capita share of water in Jordan is projected to drop to 90m$^3$/year in 2025 [6, 7]. The world water scarcity line is at 1,000m$^3$/year putting Jordan in *'absolute water shortage'* with serious consequences [4], [8]-[10].

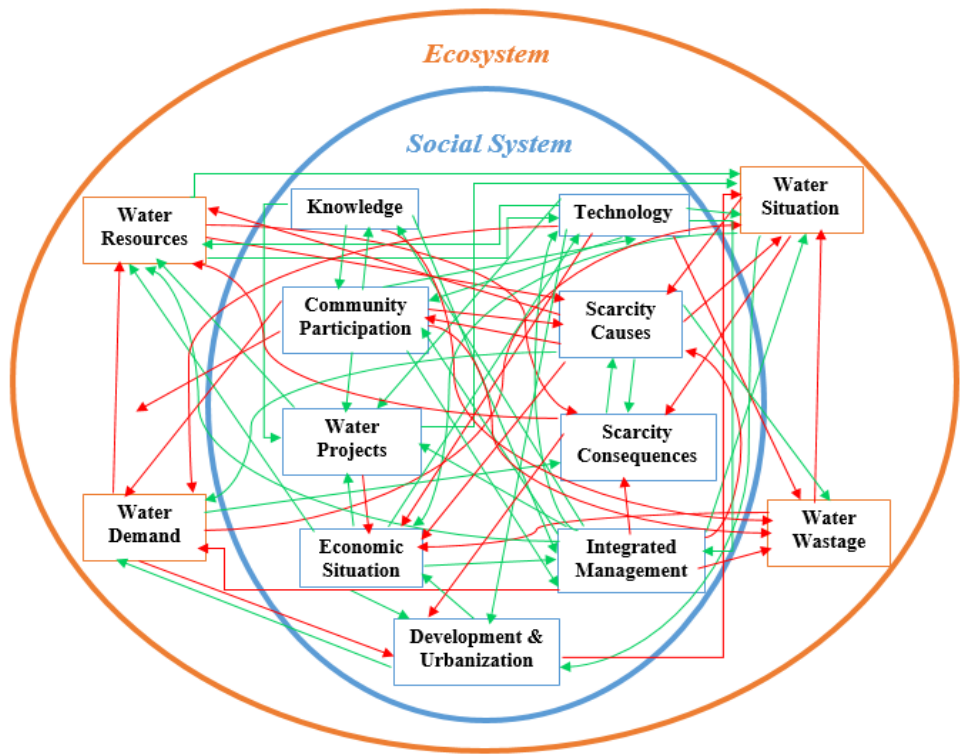

**Fig. 1** A socio-ecological system representing how its components are highly interconnected and interdependent.

This *Water Scarcity Problem* affecting a whole country is a strong representative of a large and complex socio-ecological system where the social and ecological dimensions with various positive and negative interactions create feedback loops causing nonlinear dynamical system behavior. Interactions of the social system with the ecosystem represent human actions, which in the case of water include water projects, management, pollution, urbanization etc.; whereas, reverse represent ecosystem responses to human actions such as amendments in water situation, resources and demand.

Advanced soft computing approaches such as artificial neural networks (ANN), fuzzy logic (FL), genetic algorithm (GA), and neuro-fuzzy (NF) systems [11]-[14] have been proven their effectiveness in modelling complex dynamical systems, for example, modelling the problems of water resources management.. Researchers in [14] used ANN and multiple linear regression models to assist decision-makers manage reservoir water quality. These methods are challenged when dealing with human perceptions. Adaptive Neuro-Fuzzy Inference Systems (ANFIS), ANN, and Gene Expression Programming (GEP) were compared to estimate daily river sediment volume in [11] revealing that the most accurate method was GEP. Tayfur [12] compared soft computing approaches such as ANN, FL and GA with hard computing methods for addressing specific water resources engineering issues and found soft computing methods more appropriate for modeling them. An example of using what-if scenario simulations for water resources planning and management was the integrated framework in [15] where a river-operation model is integrated with a hydrologic model to simulate associated use of surface and groundwater. Policy scenario evaluations revealed that groundwater sustainability could be achieved from new/altered management regimes. Another example of scenario simulations on a dynamical system is in [16] for recommending solutions for desertification in Ordos, China. Most of these approaches rely on complex algorithms/methods to identify crucial factors/relationships and capture nonlinear system dynamics. Moreover, they might not be appropriate in areas of conflict. Thus, an approach that is able to mimic human thought expressing their experiences, explore ill-defined factors and relationships and deal with ambiguity and uncertainty inherent in human reasoning, especially in relation to large complex systems, is in need.

Kosko introduced Fuzzy Cognitive Map (FCM) [17] to represent vagueness in socio-ecological systems and simulate systems dynamics within an Auto-Associative Neural Network (AANN). FCM is a bi-directional graph with feedback loops that capture nonlinear systems dynamics. It represents domain knowledge in relevant variables linked through directed imprecise cause-effect relationships as shown in Fig. 2 for an example FCM drawn by a stakeholder depicting their perceptions on water scarcity. FCM can represent a stakeholder perception regardless of his/her level of knowledge. FCMs representing different perceptions can be then combined into stakeholder group or whole collective perception to gain a comprehensive understanding of the problem from individual, group, and whole collective perspectives. FCM can be simplified (condensed) into a smaller number of nodes and connections for clearer understanding and gaining meaningful insights. Finally, FCM allows the simulation of what-if scenarios that would lead the system to an overall better state than the current state.

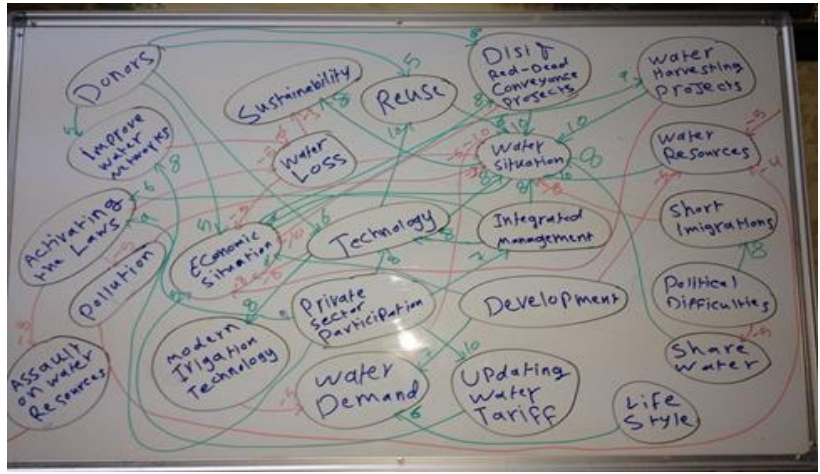

**Fig. 2** An FCM represents an view of a stakeholder on water scarcity problem. Green and red arrows indicate positive and negative influences, respectively. Numbers on arrows are the degree of the influences (connection strengths or weights).

Since Kosko [17] presented FCM, researchers have used it for modelling and managing diverse complex systems [18]-[23]. The authors in [24] used FCM to analyze conflicts in stakeholder perspectives on water use and policy regarding to a river basin in Greece, and in trans-boundary rivers in a basin shared between Greece, Bulgaria, and Turkey. Ozesmi and Ozesmi applied a multi-step FCM approach to study the conflicts and desires in different perspectives of stakeholder on some Turkey's bodies of water and lakes. [25]. Another study used a participatory FCM to analyze the future impact of climate change on water [26] and the perceptions of stakeholders showed that climate change leads to a decrease in water resources and uses.

Above attempts have revealed the usefulness of FCM in modelling and addressing such complex problems; however, these FCM approaches lack rigor without: a) an appropriate representation method that can deal with different formats of imprecise inputs (i.e., fuzzy numeric and linguistic values for connection strengths of causal relationships), b) a fuzzy persuasive FCM aggregation method that takes into consideration the differences in perceptions among stakeholders to obtain a more realistic group consensus, c) an effective condensation method, for example multi-level condensation of FCMs to reach a manageable and adequate system representation, d) an effective method to reformulate the connection weights between condensed (group) nodes resulting from each level of condensation without loss of information, and e) a sound approach to policy scenario simulations and f) an appropriate method to evaluate the feasibility of the generated policies. In our previous work [27], we presented a novel semi-quantitative FCM framework addressing all the above limitations for solving complex real world problems with heightened degree of realism.

The goal of this research is to implement our FCM framework with further enhancements to develop comprehensive solutions to a complex real life participatory problem dominated by uncertainty and ambiguity - *water scarcity crisis in Jordan* [28]. Through this application, we demonstrate the main aspects/strengths of the model, provide practical solutions to the water crisis in Jordan through an in-depth study and FCM application leading up to policy simulations to select the most effective policies to mitigate the problem, and propose the model as a generic approach to solve a range of complex real life problems. The framework is a participatory decision making system based on domain intelligence in the form of stakeholder perceptions.

To present the achievement of the goal, this paper is organized as follows: Section 2 presents an overview of the water crisis in Jordan as the context for FCM. A brief overview of the semi-quantitative FCM model used for addressing this complex problem is presented in Section 3. The outcomes from analysing the interviews and development and processing of the FCMs through multi-level map condensation and map aggregation are presented in Section 4. Section 5 presents FCM policy scenario simulations for generating recommendations for mitigating the water scarcity. The proposed "*Appropriateness*" criterion for investigating the most feasible policies is demonstrated in Section 6. Finally, Section 7 provides the summary and conclusions of the study and future directions.

## 2. Water crisis in Jordan – context for FCM

Current water demand in Jordan has reached a crisis level for several reasons. The most significant is the unpredictable population growth mainly due to migration caused by regional conflicts [6, 29]. Other contributing factors are improvements in lifestyle, tourism, industrial sectors, urbanization accelerated, and advances in agriculture [6, 30]. What underpins the crisis mostly is the extremely limited surface and groundwater resources (see Table S1 in Supplementary file) that have declined due to decreasing rainfall due to climate change [31]. This has created a chronic "*Water Scarcity Problem*" [32] that must be met by a systematic and comprehensive approach involving all representative stakeholders, for example, public, managers, farmers, water experts etc. [6]. The largest water consumption in Jordan is for agriculture use (64%); the rest is for municipal use (30%), industrial (5%) and finally, tourism (1%) uses [6]. The current water availability (1015 MCM (million cubic meters)/year) must be drastically increased to meet the projected demand of 1637 MCM/year in 2022.

In an attempt to address the crisis, Jordan Government has established a strategy called *'Jordan Water Strategy 2008–2022'* forv managing demand, conserving and developing existing water resources and searching for new resources [6]. One of its main objectives is to establish two *mega water conveyance projects*- extracting groundwater (≈US$600m) and seawater (≈US$10b) - but these are very costly. Another objective is to enhance water ruse for irrigation and industry by enhancing the treatment of wastewater and greywater [33] and water desalination using modern technologies [6] but the cost of desalination is double the revenue [34]. With these strategies, the total projected 1632 MCM/year of water available in 2022 will just meet the demand with a 5 MCM deficit [6]. However, implementation of the Water Strategy is rift with challenges [35].

The research community has also been actively seeking solutions to the problem of limited water resources and water deficit. Al-Kharabsheh and Ta'any [30] recommended gathering rainwater by water harvesting ways, such as dams, water pits, and artificial groundwater recharge. They also recommended the use of modern technologies to treat wastewater. Other useful recommendation is to establish a new By-Law in order to protect groundwater from excessive using and pollution [36], use of drip irrigation technology in agriculture [34], adoption of effective water demand programs [37], and the organization of priorities among water sectors [38]. Researchers in [35] introduced a group of recommendations to the Government including: Reducing water losses by rehabilitating water networks enhancing awareness of water importance, encouraging and helping farmers to transfer to low water consumption crops and high revenue, finally enforcing laws to prevent illegal water uses.

As a problem of this magnitude affects a whole country and its people, it is crucial to incorporate the views, concerns, perceptions etc. of all stakeholders through participation, assimilate them into a consensus view to obtain a solution that is satisfactory overall. However, a challenge for this has been the lack of a well-developed participative approach that collects different perspectives from different stakeholders, such as: policy makers, experts, private sector, and public stakeholders, and reflect them in a transparent way in the final policies derived. In this research, we implement our FCM framework developed to address these very specific concerns.

## 3. Modelling water scarcity as a complex system using Fuzzy Cognitive Maps (Methodology)

We briefly describe the semi-quantitative FCM model that we proposed in [27] and the new enhancement made in the current study to put the "*Water Scarcity problem*" in the context of FCM. The model development flow chart [27] is in the Supplemental file (Fig. S1). It consists of 8 integrated steps: 1) Interviews are conducted with relevant stakeholders; 2) The data collected from interviews are transformed into FCMs. 3) An additional recommended step is to review the interviews and FCMs to find any inadvertently omitted data; The next steps convert FCMs and subsequent processing into a fuzzy 2-tuple format: 4) FCMs are represented in fuzzy 2-tuple which was an advancement designed to unify different imprecise data formats (numeric and linguistic) into a single framework; 5) Large and complex FCMs are condensed into smaller and simpler representative maps at several levels based on FCM node credibility weights; Here, a number of advancements have been made to identify influential nodes and assign credibility weights to nodes for computing new weights after condensation to preserve the relative influence of nodes in the condensed FCMs; 6) A fuzzy weighted FCM aggregation process is used to obtain group FCMs representing all stakeholder groups and social (collective) FCM representing all stakeholders. This step benefits from a new approach to assigning credibility weights to FCMs to properly reflect the relative merits of knowledge of different stakeholders. 7) The resulting individual, group and social FCMs are analyzed by graph theoretic and statistical measures to gain a deeper understanding of the problem which leads to: 8) FCM simulation of selected what-if policy scenarios based on influential nodes. In this study, we propose a new addition to the framework- a fuzzy *Appropriateness* criterion- to evaluate candidate policy solutions according to their feasibility for implementation. We briefly describe the model steps and demonstrate their application on the data from the case study.

### 3.1 Water Scarcity in the context of FCM Model

This study included 35 individual face-to-face semi-structured qualitative interviews based on a questionnaire that included many open-ended questions covering causes and consequences of water scarcity and potential solutions from the perspective of individual participants [28]. The flexibility of the semi-structured interviews allowed participants to explore the issues freely. Table S2 shows the questionnaire prepared for the interviews. A total of 35 participants were randomly selected from the following stakeholder groups from the regions mostly affected by water scarcity: private sector (6 participants), householders (public) (8), experts (7), managers (7), and farmers (7). The first part of the interview allowed participants to explore and reflect on the water scarcity issue openly and freely. In the next part of the interview, set at a later date to give participant time to reflect, the information generated from the interviews and any new reflections were converted into variables and positively or negatively causal relationships between them to determine the system behaviour and transferred to FCMs by the participants. The average time (± standard deviation) for an interview was 89.05 ± 21.39 minutes (from 60 to 144 minutes) and for FCM reviewing and developing was 149.14 ± 31.09 minutes (from 100 to 240 minutes). The interviews produced a wide range of variables that affect water scarcity and potential solutions. Fig. 3 shows that the number of new variables added by each new FCM quickly diminishes in subsequent interviews indicating that the selected sample size has ensured comprehensiveness of the domain data collected. It also indicates a great deal of awareness, familiarity and understanding of the water scarcity issue across population represented by this spectrum of randomly selected stakeholders.

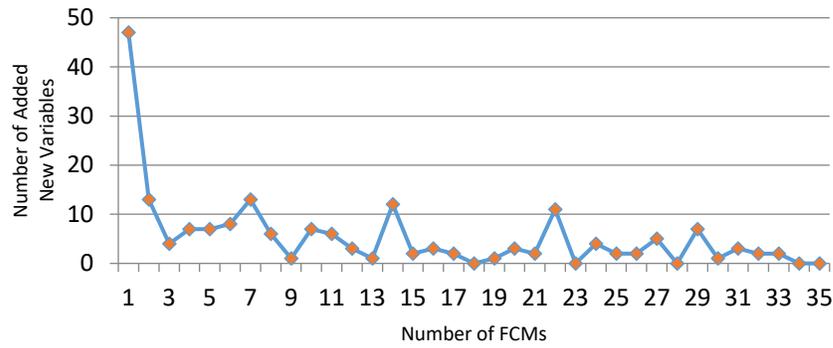

**Fig. 3** New variable vs the number of FCMs

In total, there were 186 variables represented in the FCMs (Table S8 -last column). They represented broad categories covering causes of water scarcity and potential solutions. Prominent causes included limited water resources, demand, wastage, development, over population and limited funding. Among prominent solutions were law enforcement, community participation, new technology, and new and enhanced water projects. These causes, consequences and solutions were shown to interact in complex ways as shown in Fig. 1. In summary, all participants were acutely aware of water scarcity and its causes and potential solutions. They are in agreement in thinking that water scarcity is caused by many factors including population increase leading to increased demand, lack of surface and ground water resources, lack of rainfall, wastage, leakage in water pipe networks, fragmented management efforts by the authorities, illegal use of water and weak law enforcement and inadequate funding. They saw that the solution to the crisis is in developing new mega water projects (sea water and groundwater conveyance) incurring huge costs, strengthening and expanding existing water storage and rain harvesting systems, investing in new technology for wastewater treatment and desalination, integrated management, community participation and stronger law enforcement and more. In this study, we evaluate these options for their efficacy and feasibility using the proposed FCM framework.

From the 35 conducted interviews, 35 FCMs were developed by the stakeholders. Fig. 2 is an example FCM developed by a stakeholder. Here factors/variables are depicted in circles and their relations by arrows (red-inhibitory; green- reinforcing) with a number indicating the strength of the relationship (weight). The stakeholders preferred values of numeric and linguistic to represent weights in FCMs using either or [-10, 10] or [-1, 1] intervals for numerical values or 13 or 11 linguistic expressions. This gave participants greater flexibility to describe the degree of influence between the variables based on their skill level or perception. The 35 FCMs were reviewed and updated which added new variables and relationships among them to the FCMs that were missing in the original FCMs.

We used a 2-tuple fuzzy linguistic representation model to represent and unify the two different imprecise input formats (linguistic and numeric) for the connections strengths or weights defined by stakeholders [39]. This representation is very effective for processing in all processes of FCM. It uses a pair of symbolic fuzzy values (2-tuple) to represent each of the imprecise numeric or linguistic values. This avoids loss of information and keep consistency throughout all subsequent computations. This way both numeric and linguistic connection weights are converted to one format of 2-tuple. For the purpose of calculations, it converts symbolic fuzzy values into numeric fuzzy values ($\beta$ values), and any $\beta$ value can be converted back to a 2-tuple. Finally, it is able to convert the fuzzy $\beta$ values into crisp values, and vice versa. Brief overview of the approach is presented here and the reader is referred to [27, 28] for details.

| | | | |
|---|---|---|---|
| $s_{-6}$ | *negatively very very high (-VVH)* | $s_1$ | *positively very low (VL)* |
| $s_{-5}$ | *negatively very high (-VH)* | $s_2$ | *positively low (L)* |
| $s_{-4}$ | *negatively high (-H)* | $s_3$ | *positively medium (M)* |
| $s_{-3}$ | *negatively medium (-M)* | $s_4$ | *positively high (H)* |
| $s_{-2}$ | *negatively low (-L)* | $s_5$ | *positively very high (VH)* |
| $s_{-1}$ | *negatively very low (-VL)* | $s_6$ | *positively very very high (VVH)* |
| $s_0$ | *zero (Null)* | | |

Fig. 4 The linguistic terms used by stakeholders.

The 2-tuple model is based on a symbolic linguistic model proposed in [40]. It represents linguistic information by an ordered linguistic term set, $S = \{s_0, \ldots, s_g\}$, where the number of linguistic terms/fuzzy sets in $S$ is equal $g+1$. In our study, two linguistic term sets, $S_1 = \{s_{-6}, \ldots, s_6\}$ (13 fuzzy sets) and $S_2 = \{s_{-5}, \ldots, s_5\}$ (11 fuzzy sets), were used to represent the two formats used by stakeholders in 2-tuple and $\beta$ values, as shown in Fig. 4. These two linguistic sets were normalized into one standard set called Base Linguistic Term Set (*BLTS*) using the method in [41]. We considered the 13 fuzzy subsets as the *BLTS* and normalized the weights represented by 11 fuzzy subsets into values in *BLTS*. Triangular Fuzzy sets were used in both cases and Fig. 5 shows the 13 fuzzy subsets with membership function $\mu$ which for a fuzzy subset $A$ is defined by three parameters $(a, b, c)$ where $a \leq b \leq c$

in the universe of discourse $X$ and is calculated as follows:

$$\mu_A(x) = \begin{cases} x - a/b - a, & a \leq x \leq b \\ c - x/c - b & b < x \leq c \\ 0 & \text{otherwise} \end{cases} \quad (1)$$

Here the universe of discourse refers to the weight between any two nodes in the FCM. In the case of a numeric imprecise value ($x$), the membership function of value $x$ associated with each $s_i$ (i.e., $\mu_{si}(x)$) is calculated using Eq. 1. Then, we calculate a $\beta$ value as the numeric result of a symbolic aggregation of membership functions over labels $i$ assessed in $S$ and obtained from:

$$\beta = \sum_{i=-g}^{g} i \cdot \mu_{si} / \sum_{i=-g}^{g} \mu_{si} \quad (2)$$

where $g$ is the number of linguistic sets.

Now we describe the generation of fuzzy 2-tuples ($s_i$, $\alpha$). Let $\alpha = \beta - i$ be two values, such that $i \in [0, g]$ and $\alpha \in [-0.5, 0.5)$; and $i = round(\beta)$, where round (.) is the mathematical rounding operation; then $\alpha$ is called a symbolic translation [38]. Using this definition, the 2-tuple ($s_i$, $\alpha$) equivalent of $\beta$, $\Delta(\beta)$, is calculated as:

$$\Delta(\beta) = \begin{cases} s_i, & i = round(\beta) \\ \alpha = \beta - i, & \alpha \in [-0.5, 0.5) \end{cases} \quad (3)$$

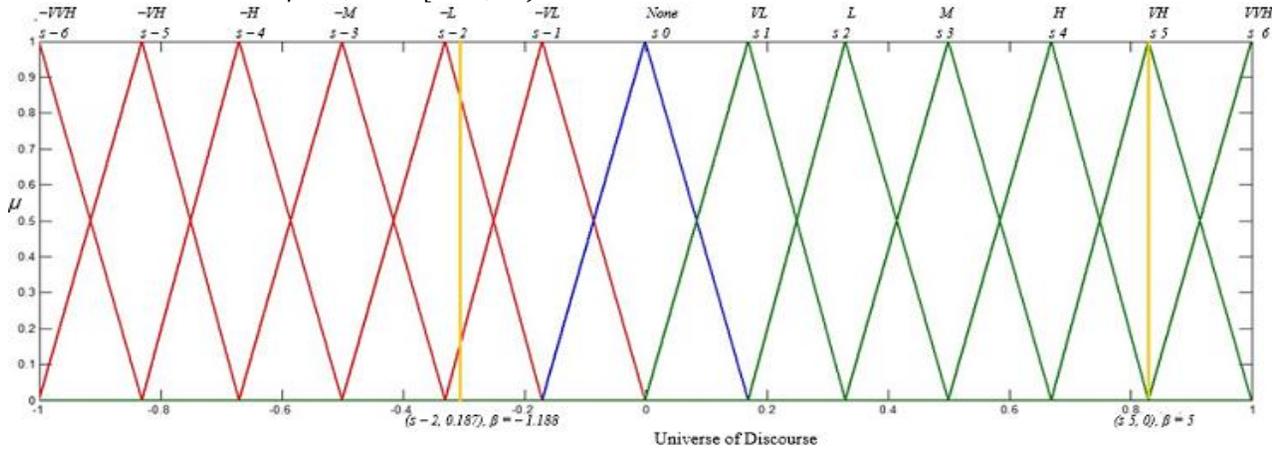

**Fig. 5** The 13 linguistic symbolic terms represented by 13 triangular membership functions of [27] and two examples of conversion of imprecise input data into fuzzy 2-tuple.

And the equivalent numerical value ($\beta$) to 2-tuple ($s_i$, $\alpha$), $\Delta^{-1}(s_i, \alpha)$, is calculated as:

$$\Delta^{-1}(s_i, \alpha) = i + \alpha = \beta \quad (4)$$

The equivalent 2-tuple representation of $s_i$ is obtained by adding a value of 0 as the symbolic translation:

$$\Delta(s_i) = (s_i, 0) \quad (5)$$

A 2-tuple value of a linguistic weight represented by a linguistic term $s_i$ in $S$ is directly obtained using Eq. 5, and its equivalent $\beta$ value is calculated using Eq. 4. In the case of a weight being represented by a numeric imprecise value in the range $[-1, 1]$, first its $\beta$ value is calculated from Eq. 2 and then its equivalent 2-tuple value is obtained from Eq. 3. Fig.5 shows one example each of linguistically and numerically expressed weights with yellow vertical lines on the right and left hand sides, respectively. In the former case, the expressed weight is *VH* or fuzzy subset $s_5$; therefore, $i=5$. From Eq. 5, corresponding 2-tuple is ($s_5$,0) and from Eq.4, $\beta=5$. In the latter case, numeric weight is -1.813 and belongs to both –*VL* ($s_{-1}$) and –*L* ($s_{-2}$) with corresponding membership function values given by Eq. 1. Then from Eq. 2, $\beta= -1.813$. From Eq. 3, $i=-2$, $\alpha=0.187$ thus 2-tuple is ($s_{-2}$, 0.187). Table 1 presents the original (imprecise numeric) weight matrix entries for few selected nodes in the example FCM in Fig. 6 and their corresponding fuzzy $\beta$ values (shown within brackets) (refer to Tables S3 and S4 for all entries). This way we represented all 35 FCMs in the unified (2-tuple) format and converted weights to fuzzy numeric $\beta$ values for subsequent computations.

**Table 1** The original weight matrix entries for few selected nodes in FCM in Fig. 4 and corresponding fuzzy $\beta$ values (within brackets)

|  | $c_1$ | $c_2$ | $c_4$ | $c_6$ | $c_{12}$ |
|---|---|---|---|---|---|
| $c_1$ | 0 | 0 | 0.37 (2.21) | 0 | -0.71 (-4.27) |
| $c_3$ | -0.71 (-4.27) | -0.13 (-0.77) | 0 | 0 | 0.26 (1.57) |
| $c_5$ | -0.63 (-3.76) | 0 | -0.34 (-2.07) | 0 | 0 |
| $c_8$ | -0.1 (-0.61) | 0 | 0.48 (2.88) | 0 | 0 |
| $c_9$ | 0.48 (2.89) | 0.27 (1.63) | -0.47 (-2.80) | 0.25 (1.47) | 0 |
| $c_{13}$ | -0.17 (-1.01) | -0.63 (-3.78) | -0.34 (-2.06) | 0 | 0.42 (2.51) |

After the fuzzy representation of the FCMs, they were condensed into smaller representative maps. We used two levels of a semi-quantitative condensation resulting in 3 levels of FCMs: 1) *original variables*, *i.e.*, variables originally defined by a stakeholder; 2) *key variables*, *i.e.*, variables resulting from the 1st condensation; and 3) *concepts*, *i.e.*, variables resulting from the 2nd condensation. In each level, similar variables were subjectively clustered based on their belonging to the same category. These categories are groups of key variables or concepts, respectively, at the higher level and similar variables clustered are those at the lower level (variables or key variables, respectively). At the first level, the 186 original variables were condensed into 42 key variables which were then condensed into 13 concepts at the second level.

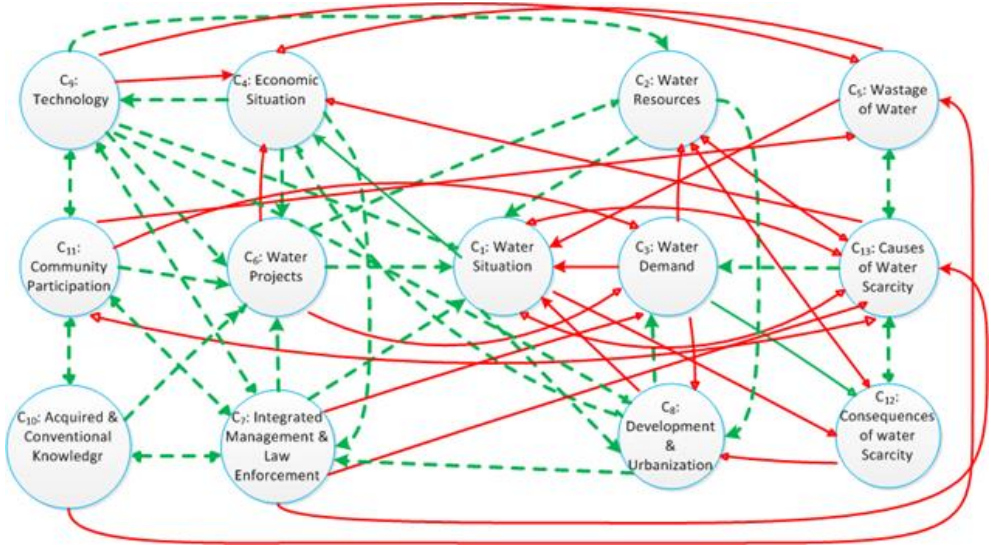

**Fig. 6** An example FCM representing variables and interactions as they influence water scarcity. For clarity, arrow weights are not shown on the graph.

To calculate the new weights between groups at the higher level, we used the method we proposed in [27] where these values are represented by fuzzy numeric $\beta$ values to avoid loss of information. Credibility weights (*CW*) assigned to nodes at the lower level are used to transfer the nodes and their connections to a higher level. These *CW* were obtained based on a novel consensus centrality measure (CCM) proposed by us in [42] (Eq. 6-9). The CCM of a node reflects its importance in an FCM; *i.e.*, a node with a higher CCM value is of higher importance and more central. The CCM is calculated using 2-tuple fuzzy representation model and based on three common centrality measures from social networks theory - *degree*, *closeness* and *betweenness*. *Degree* centrality measures the strength of interaction of a node with other nodes, *i.e.*, sum of incoming and out-going connection strengths [43, 44] (Eq.6 part 1). *Closeness* centrality measures how quickly a node is reached from other nodes [44, 45] (Eq.6 part 2). *Betweeness* centrality measures how frequently a node is on the path of communication between other nodes [44] (Eq.6 part 3):

$$Cen_D(C_i) = \sum_{j=1}^{N} |w_{ij}|;$$

$$Cen_C(c_i) = 1/\sum_{t=1}^{N} d_G(t, c_i); \qquad (6)$$

$$Cen_B(c_i) = \sum_{s,t=1}^{N} \sigma_{st}(c_i)/\sigma_{st}$$

where $Cen_D(c_i)$, $Cen_C(c_i)$ and $Cen_B(c_i)$ are the degree, closeness and betweenness centrality of the node $c_i$ and $N$ is the number of nodes connected to node $c_i$ in FCM. The $w_{ji}$ is the weight of the connection entering (or leaving) node $c_i$ from (or to) node $c_j$; $d_G(t, c_i)$ ($t \neq c_i$) is the shortest path from node $t$ to node $c_i$ and $\sigma_{st}$ is the number of shortest paths from node $s$ to node $t$, and $\sigma_{st}(c_i)$ is the number of shortest paths from node $s$ to node $t$ that pass through node $c_i$.

The CCM for a node $Cen_{Cons}(c_i)$ is a weighted average of the above measures (Eq.7):

$$Cen_{Cons}(c_i) = b_D \cdot Cen_D(c_i) + b_C \cdot Cen_C(c_i) + b_B \cdot Cen_B(c_i) \qquad (7)$$

where $b_D$, $b_C$ and $b_B$ are the respective prioritization weights; $b_D + b_C + b_B = 1$.

The three centrality measures for FCMs are calculated from the corresponding node centralities as

$$Cen_D(FCM) = \sum_{i=1}^{N} \frac{(Cen_D^* - Cen_D(c_i))}{(N-1)};$$

$$Cen_C(FCM) = \sum_{i=1}^{N} \frac{(Cen_C^* - Cen_C(c_i))}{(N-1)(N-2)(N-3)}; \qquad (8)$$

$$Cen_B(FCM) = \sum_{i=1}^{N} \frac{(Cen_B^* - Cen_B(c_i))}{(N-1)}$$

where $Cen_D(FCM)$, $Cen_C(FCM)$ and $Cen_B(FCM)$ are the degree, closeness and betweenness centrality of FCM, $Cen^*_D$ is the maximum degree, $Cen^*_C$ is the maximum closeness and $Cen^*_B$ is the maximum betweenness centrality of nodes in the FCM. $Cen_D(c_i)$, $Cen_C(c_i)$ and $Cen_B(c_i)$ are the corresponding centrality values of any node $c_i$. The consensus centrality measure of FCM $Cen_{Cons}(FCM)$ is given by

$$Cen_{Cons}(FCM) = b_d . Cen_D(FCM) + b_C . Cen_C(FCM) + b_B . Cen_B(FCM) \quad (9)$$

**Table 2** Degree, Closeness, Betweenness and Consensus Centrality values and Credibility weights of nodes in Table 1

| Node | $Cen_D$ | $Cen_C$ | $Cen_B$ | CCM | CW |
|---|---|---|---|---|---|
| $c_1$ | 5.66 | 4.12 | 1.27 | 3.69 | 0.113 |
| $c_3$ | 3.06 | 0.71 | 0.38 | 1.38 | 0.042 |
| $c_5$ | 1.56 | 0.75 | 0.5 | 0.94 | 0.029 |
| $c_8$ | 2.14 | 3.6 | 1.13 | 2.29 | 0.07 |
| $c_9$ | 3.93 | 3.17 | 0 | 2.37 | 0.072 |
| $c_{13}$ | 3.71 | 6 | 6 | 5.24 | 0.159 |

In this study, prioritisation weights were subjectively assigned, which is the same for all measures. Table 2 (columns 2-5) shows the above measures in $\beta$ values for the same nodes from FCM in Fig. 6 shown in Table 1 (refer to Table S5 for all entries).

To calculate *CW* of nodes, their CCM in $\beta$ values are first defuzzified into crisp numeric values in the interval [0, 1] using Eq. 10 and 11 [40]. First $\beta$ values are transformed into fuzzy two tuple $(s_i, \alpha)$ using function $\delta$:

$$\delta(\beta) = \{(s_h, 1-\gamma), (s_{h+1}, \gamma)\} \quad (10)$$

where $h = trunc(\beta)$, $\gamma = \beta - h$, and $S_h$ and $S_{h+1}$ are linguistic terms in $S$, in which $\beta$ has membership. Then the two 2-tuples are converted into a crisp value in the interval [0, 1] using function $\kappa$

$$K\big((s_h, 1-\gamma), (s_{h+1}, \gamma)\big) = CV(s_h) * (1-\gamma) + CV(s_{h+1}) * \gamma \quad (11)$$

where $CV(.)$ is a function with characteristics of defuzzification techniques such as *Mean of Maximum (MoM)*. Then, for each node $c_i$, its fuzzy numeric CCM value is normalised by the sum of CCM values of all nodes in the FCM to obtain its credibility weight $(cw_i)$ (Table 2, Column 6).

Based on the above, the process of calculating new weights between groups at the higher level is as follow: a) initialize to zero values the connection weights between groups at the higher level, b) for the connection between two groups (*i.e.*, $g_i$ and $g_j$), only the non-zero valued connections between nodes in these groups were identified at the lower level. For the identified nodes in each group, we use their *CW* at the lower level to calculate a new *CW* for them using:

$$New\_cw_{ni} = cw_{ni} / sum(cw_{Ni}) \quad (12)$$

where $New\_cw_{ni}$ is the new *CW* of a selected node $n$ in $g_i$, $n = 1 \ldots N$, $N$ is the total number of selected nodes in $g_i$, $cw_{ni}$ is the lower level *CW* of the selected node $n$ in $g_i$, and $sum(cw_{Ni})$ is the sum of *CWs* at the lower level of all selected nodes in group $g_i$.

Then, Eq.13 uses these new *CW* of nodes and their connection weights at the lower level to calculate a condensed connection weight between the two groups at the higher level:

$$g_{ij} = \sum(New\_cw_{ni} * New_{cw_{nj}} * w_{ij}) \quad (13)$$

where the right hand side depicts the sum of all connection weights $w_{ij}$ at the lower level between nodes in $g_i$ and nodes in $g_j$ after weighting each weight between two nodes by their new credibility weights assigned using Eq. 12. These steps are repeated until all connections between all groups at the higher level are calculated. The process is repeated at the next level of condensation and so on.

Once FCMs were condensed, they were combined into group and social FCMs to represent consensus perception of individual groups and that of the combined whole, respectively. First, we aggregated the FCMs of each of the five groups at all levels; *original variables*, *key variables* and *concepts* (resulting in 15 FCMs). We then obtained, for each level, a social FCM that includes the perceptions of all stakeholders (resulting in additional 3 FCMs). To do this, we used the FCM fuzzy aggregation method presented in our previous work [46]. It takes into consideration the different levels of knowledge and experience of the participants who develop FCMs by weighting their FCMs according to an FCM credibility weight (*CW*). The credibility weight of an FCM is its

CCM (consensus centrality measure, $Cen_{Cons}$(FCM) obtained from Eq. 9) normalised with respect to the total CCM of all FCMs. The process of aggregation starts by initializing a zero weight matrix for the group/social FCM (*Soc*) (Eq. 14) to include nodes and connection weights in all individual/group FCMs that could result from combining the corresponding FCMs into Group/social FCM. Then, for each $FCM_k$ in the individual/group, we recall its connection weights in *β* values, FCM credibility weight $cw_k$, and number of nodes. Then, the weight matrix of each $FCM_k$ is augmented to include all nodes in all individual/group FCMs. The column and row of each new node added to the matrix are filled with zero values. The augmented matrix is then multiplied by the FCM credibility weight $cw_k$ to obtain a weighted augmented map *($FCM_{wk}$)* (Eq.14 left part) which is then aggregated into the weight matrix of group/social FCM (Eq.14 right part):

$$FCM_{wk} = FCM_K * cw_k; \qquad Soc = Soc + FCM_{wk} \qquad (14)$$

At this stage, a large amount of data in the form of 123 FCMs have been generated including individual FCMs in the form of original and condensed variables (key variables and concepts) (35*3) as well as group and social maps at the three levels of condensation (3*5+ 3) containing rich information for analysis. We will introduce the new method to determine *Appropriateness* of policy solutions at the simulation stage.

## 4. Analysis of FCMs of the Case Study

We conducted a graph theoretic and statistical analysis of FCMs to gain a deeper understanding of the problem [28]. For all 123 FCMs, we determined the number of variables and connections, density and map and node CCM values. Density is connectedness with respect to maximum possible connectivity in an FCM; density of 1.0 indicates a fully connected map. From the CCM of maps and their nodes we determined most influential maps (*i.e.,* stakeholders and groups) and nodes (*i.e.,* variables, key variables and concepts) with strong influence on water scarcity. By systematically analysing the individual, group and social FCMs at the *variable*, *key variable* and *concept* levels we found that the most important '*variables*' inform the most important '*key variables*' which in turn inform the most important '*concepts*' (see Table S8) indicating with a high level of certainty that the maps at each level have captured the essential features to accurately represent the overall perception of the stakeholders.

Table 3 shows some results for the three *social* FCMs produced by aggregating all 35 FCMs at the levels of *variables* (V), *key variables* (KV) and *concepts* (C). It shows that overall V, KV and C are broadly in agreement regarding factors that influence water scarcity the most; all 3 levels have captured economic impact, water demand, availability of water resources and sustainability, water loss/wastage as their top influencers. At the level of variables, social FCM had 186 variables, 2682 connections and 0.078 density. Of note, the important variables in the social FCM according to their CCM values were similar to the important variables in the individual and group FCMs at that level (not shown). At the level of key variables, social FCM had 42 key variables, 771 connections and 0.448 density. The top two key variables *'National Funding'* for water related matters and *'Water Situation'* also corresponded to the top key variables in individual and group FCMs (not shown). At the level of concepts, social FCM had 13 concepts, 135 connections and a very high density (0.865). The Concepts *Economic Situation, 'Water Situation'* and available *'Water Resources'* have retained their prominence in the social map (Table 3). (The 13 concepts were: *'Water Situation (A)', 'Water Resources (B)', 'Water Demand (C)', 'Economic Situation (D)', 'Wastage of Water (E)', 'Water Projects (F)', 'Integrated Management and Laws (G)', 'Development and Urbanization (H)', 'Technology (I)', 'Acquired and Conventional Knowledge (J)'* of people about water, *'Community Participation (K)', 'Consequences of Water Scarcity (L)',* and *'Causes of Water Scarcity (M)'*). These concepts summarily capture the causes and consequences of water scarcity contained in the original 186 variables. Basically, maps at all three levels highlight the negative impact of the economic situation, demand, wastage and development and urbanisation on the water scarcity crisis in Jordan and the positive solutions through technology, proper water management and laws, knowledge and public awareness of water and water situation, community participation and water projects (new and improved).

The analysis of the five group FCMs at the concept level revealed similar properties across the groups (Table S6). Their most important concepts were: *'Causes of Water Scarcity (M)', 'Development and Urbanization (H)', 'Integrated Management and Laws (G)', 'Economic Situation (D)', 'Water Situation (A)', 'Technology (I)', 'Water Resources (B)'* and *'Water Demand (C)'*. Similarity of CCM values have caused some concepts to shift but these top concepts overlap with those for the social map (Table 3) and individual maps (not shown) at this level. The above results and overall FCM analysis revealed that stakeholders in all 5 groups were generally united in their perception of the water problem, its causes and consequences and potential solutions and obstacles. This is evident from the similarity of their FCMs and influential variables. Thus the condensed maps have retained the general character of these perceptions and provided a clearer overall view at each level. Next Section presents the FCM simulations to determine effective solutions for mitigating water scarcity, for more information please refer to [28].

**Table 3** Some graph theory indices and top 5 nodes in social FCMs at original variable (V), key variable (KV) and concept (C) levels

| Level | No of V, KV,C | No of Cons. | FCM Density | 1st V, KV,C | 2nd V, KV,C | 3rd V, KV,C | 4th V, KV,C | 5th V, KV,C |
|---|---|---|---|---|---|---|---|---|
| V | 186 | 2682 | 0.078 | DA1 | AA1 | EA1 | AC1 | HA2 |
| KV | 42 | 771 | 0.448 | DA | AA | GD | AC | EA |
| C | 13 | 135 | 0.865 | E | C | D | A | B |

DA1: Financial Situation, AA1: Quantity of Water, EA1: Technical Loss & Leaking of Water Networks, AC1: Sustainability of Water, HA2: Agricultural Development, DA: National Funding, AA: Water Situation, GD: Management of Demand & Supply, AC: Sustainability of Water, EA: Loss of Water Networks, E: Wastage of Water, C: Water Demand, D: Economic Situation, A: Water Situation, B: Water Resources.

## 5. Policy scenario simulations on FCM

An additional value of the CCM measure is that it can help identify the most important (central) nodes in an FCM and the most important FCMs in the whole set of FCMs. These important nodes make a significant contribution to their FCM (system) and could be used as influential policy instruments to change the behaviour/outcome of the system in FCM simulations. For this purpose, FCM is considered as a dynamical system in the form of Auto-associative Neural Networks (AANN) where each node is a neuron [25] that feeds its output to other neurons connected to it. The nodes take values in the range [0, 1] indicating their state or level indicative of low or high. In each simulation iteration, a new state vector is calculated using Eq. 15.

$$A_i^{t+1} = f(\sum_{j=1}^{N} A_j^t * w_{ij}) \quad (15)$$

where $A_i^{t+1}$ and $A_j^t$ are the states of node $c_i$ at time $t+1$ and $t$, respectively, $N$ is the number of nodes, $w_{ij}$ is the strength of connection from node $c_i$ to node $c_j$, and $f$ is a nonlinear function (logistic function used in this study) that determines the new state of the node $c_i$ in [0, 1] based on the weighted sum of inputs it receives from other nodes (Eq.15). Starting from the initial state of nodes (inputs) the simulation process proceeds in a number of iterations until a steady state is reached.

This study suggests two phases of simulation to conduct policy simulations on the *social* FCM as follows: First, the *social* FCM at the *concept* level is simulated with the present (default) state of nodes until steady state. This can serve two purposes: as it can show the most sensitive nodes in the system, they could corroborate the policy variables deemed suitable for simulation from the stakeholder maps based on their high CCM. Further, it can serve as the base for the next stage where FCM is simulated with new policies. The policy simulation means holding one or more important nodes at fixed states in each iteration of the simulation until equilibrium (new steady state) is reached. Comparing the steady state results from the two simulations, the policies (concepts) that produce the most desirable outcome (*improved water situation*) are selected based on their highest impact on it. However, selected policies must be specified at the level of *original variables* where they are implemented. Therefore, once the most effective *concepts* have been found from the scenario simulations, the investigation follows further probing into these identified concepts to select firstly the corresponding *key variables* and then *original variables* that strongly influence the desired outcome using scenario simulations on the *social* FCM at these two levels. To do this, each *key variable* in each of the selected *concepts* is considered as a policy instrument and clamped in the social FCM at the level of *key variables* and the scenario is simulated to ascertain their influence on the key variables representing water situation. From these simulations, influential *key variables* are selected. Then, the scenario simulations are repeated at the *variable* level based on the influential *key variables* to find the influential *variables*. These policy variables are then evaluated for their *Appropriateness* for implementation and ranked to select the most effective policies to mitigate the water crisis as discussed in the next section.

### 5.1 Phase 1: Default simulations on social FCMs

In this phase, *social FCM* at the level of *concepts* are simulated with default initial states [28]. These initial states (shown within brackets) of the 13 concepts were as follows: 'Water Situation (A)' (0.33), *'Water Resources (B)'* (0.33), *'Water Demand (C)'* (1.0), *'Economic Situation (D)'* (0.33), *'Wastage of Water (E)'* (0.67), *'Water Projects (F)'* (0.5), *'Integrated Management and Laws (G)'* (0.67), *'Development and Urbanization (H)'* (0.83), *'Technology (I)'* (0.67), *'Acquired and Conventional Knowledge (J)'* (0.5), *'Community Participation (K)'* (0.83), *'Causes of Water Scarcity (M)')* (0.83). Stakeholders in their FCMs expressed two promising avenues to mitigate water scarcity: that *'Water Situation (A)'* can be improved mainly through enhancing *'Water Resources (B)'* and reducing *'Water Demand (C)'*. From now on, these three concepts are considered as the target concepts (underlined on the above list). Other concepts influence the target concepts positively or negatively. Therefore, analysis of the results of the two phases of simulation (default and policy scenarios) focuses on the concepts (inputs) that most strongly impact the target concepts (outputs). We formulated the above initial state values based on information from literature on the water situation in Jordan [6]. The information was also supported/corroborated by the data from the stakeholder interviews, particularly the managers and experts. Based on this initial state and connection weights in the social FCM, simulation phase 1 implements Eq. 15 iteratively until steady state. We also repeated the simulation on the 5 group FCMs at the concept level for comparison (not shown). The systems reached steady state in 11-13 iterations.

The analysis of the influential concepts in the default simulations indicated that development of *'Water Projects (F)'*, *'Integrated Management and Laws (G)'*, controlling *'Development and Urbanization (H)'*, especially agriculture and construction, increased use of *'I'* for water treatment etc. and increased *'Community Participation (K)'* as the most influential concepts that can contribute to *"mitigating the Water Scarcity Problem"*. These variables are also among the top concepts according to CCM in Tables 2 and S5. Therefore, these concepts were selected for policy scenario simulations; for example, raising each of them, in a separate scenario, to a high value (*i.e.*, 1.0) or low value (i.e., 0) and keeping it at that level throughout the simulation to ascertain the final system state with respect to the selected target nodes ('Water Situation (A)', 'Water Resources (B)' and 'Water Demand (C)').

Next section presents these scenarios and results.

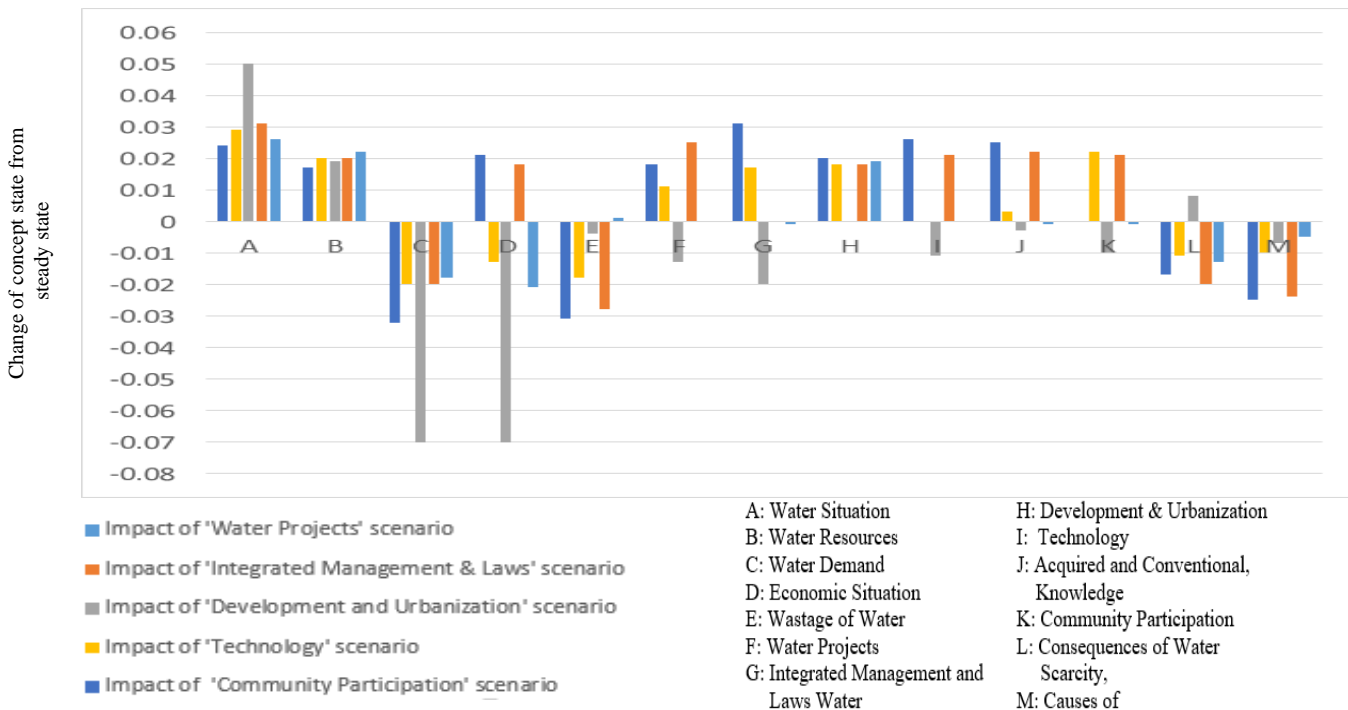

**Fig.7** Results from the five selected policy scenario simulations showing their influence on the targets and other nodes

### 5.2 Phase 2: Targeted policy scenario simulations

Simulation of each policy scenario and its outcomes are analysed in the following manner. First, the *social* FCM is simulated based on the target policy scenario until steady state. Then the steady state of nodes is compared with the corresponding state from the default simulation. The difference indicates the effect of the targeted policy [28].

Fig. 7 depicts the results from the five targeted policy scenario simulations. As shown, increasing *'Water Projects (F)'* (A-light blue bars) improves *'Water Situation (A)'*, *'Water Resources (B)'* and *'Development and Urbanization (H)'* and reduces *'Water Demand (C)'*, *'Community Participation (K)'* and to some extent *'Causes of Water Scarcity (M)')*. The major challenge for implementing this scenario is the decline in the *'Economic Situation (D)'*. When *'Integrated Management and Laws (G)'* (G-brown bars) is fixed at a high value of 1.0, the results show that it is an effective policy overall as it improves the states of all the other 12 concepts. In the simulation with restricted Development and Urbanization (H-grey bars) with a fixed value of 0, results indicates that although it increases *'Water Situation (A)'* mainly by reducing *'Water Demand (C)'*, its overall impact is poor as it negatively impacts *'Economic Situation (D)'*, implementation of *'Water Projects (F)'* and *'Economic Situation (D)'*, employment of *'Technology (I)'* and *'Community Participation (K)'*. Development and urbanization are essential aspects in any country; therefore, this scenario is impractical and inappropriate. Regarding Policy scenario four with investment in modern '*Technology (I)*' (I-yellow bars) fixed at a high level of 1.0, the results show that this scenario makes a noticeable change in all other concept states. The advantages of this policy over the high level of *'Water Projects (F)'* are: it increases the efficiency of the 'Integrated Management' and *'Community Participation (K)'*, and decreases *'Wastage of Water (E)'*. However, both scenarios worsen the *'Economic Situation (D)'*. Finally, Promoting and strengthening *K* policy scenario (K- dark blue bars) with a fixed state of 1.0 shows approximately similar results to scenario 'Integrated Management and Laws'.

Thus, all five scenarios could mitigate the "*Water Scarcity Problem*" through enhancing water situation and resources and reducing water demand and wastage to varying degrees. However, some face challenges to their adoption. For example, increasing *'Water Projects (F)'* and *'Technology (I)'* require adequate funding but this could be resolved through donors, loans and private sector involvement. However, reduction to *'Development and Urbanization (H)'* is inappropriate and hence it would not be considered as a potential solution. The other policies are promising towards generating useful recommendations. Next we performed two combined policy simulations; one with all 5 policies and the other excluding *'Development and Urbanization (H)'* (Figs. S2.A and B). Although the 1st scenario significantly improves the water situation and decreases demand, the 2nd scenario is more convincing as it does not cause a significant decline in the *'Economic Situation (D)'* (in fact there is a marginal increase) and further it shows a greater reduction in *'Wastage of Water (E)'* and increase in *'Development and Urbanization (H)'*. Therefore, we exclude *'Development and Urbanization (H)'* policy and select the other four policies for deeper investigation in the next section.

**Table 4** The three criteria proposed to assess the "*Appropriateness*" of a key or original variable

| Criterion | Sub Criteria | Description |
|---|---|---|
| 1. Importance | . Credibility weight | Importance of a KV or V based on its CCM value |
|  | . Number of times mentioned | Importance of a KV or V based on the number of times mentioned in FCMs |
| 2. Feasibility | . Influence on *Economic Situation* | Degree of acceptance of a KV or V based on its impact on reducing negative economic effects |
| 3. Influence | . Influence on *Water Situation* | Influence of a KV or V based on its impact on improving water situation |
|  | . Influence on *Water Resources* | Influence of a KV or V based on its impact on increasing water resources |
|  | . Influence on *Water Demand* | Influence of a KV or V based on its impact on decreasing water demand |

## 6. Novel "*Appropriateness*" criterion for investigating feasible of policy solutions

As stated before, the concepts in social FCM are condensed variables and therefore are not suitable for policy recommendations. Therefore, we carry out a deeper investigation of these concepts at their lower levels of condensation to determine which corresponding key variables and then the original variables are the most effective and feasible in producing the desired outcomes. Some of these candidate original variables could be inappropriate or difficult to implement in reality. To overcome this challenge, we propose an "*Appropriateness*" criterion to evaluate their feasibility of implementation [28]. It is based on three sub criteria: *influence* of a key or original variable on water situation, their economic *feasibility* and the relative *importance* of their role in the FCMs. The value of "*Appropriateness*" is the weighted aggregation of these 3 criteria (Table 4). In this investigation, the 4 acceptable policy concepts are probed to select and rank influential *key* and *original variables* (KV or V) according to their "*Appropriateness*". The *Importance* of a KV or V is the average of the normalized values of its credibility weight (*CW*) and how many times it is mentioned in the FCMs [25]. The *Feasibility* and *Influence* of a KV or V are assessed by simulating a new policy scenario in which the KV or V is fixed at a high value in the social FCM at the corresponding level and assessing the change in the states compared to their steady state values from the default simulation conducted at the same level. This process is similar to the above map simulations at the concept level. Specifically, the changes to the states of the KV or V that belong to *'Economic Situation (D)'* concept (Fig. 8) assess the *Feasibility* criterion, and the changes to the states of the KV or V that belong to target concepts of - *'Water Situation (A)'*, *'Water Resources (B)'* and *'Water Demand (C)'*- (Fig. 9) assess the *Influence* criterion [28].

To calculate the appropriateness of a *key variable* or *original variable*, first, the values of the above 3 criteria are expressed as percentages. Then, in this study, 0.25, 0.25, and 0.5 weights were used to prioritize the *Importance*, *Feasibility* and *Influence*, respectively. The largest weight is given to *Influence* because it focuses on solutions to "*Mitigating Water Scarcity*". The weight of *Feasibility* depends on the level of concern for the economic situation. For the *Feasibility* criterion, we considered the change in the steady state values of KV or V that belong to *'Economic Situation (D)'* concept, which are: *'National Funding'* at the level of KV and country's *'Financial Situation and Income'* and *'Government Budget'* at the level of V (Fig. 6). The changes to steady state are normalised, averaged, and sign changed to negative to obtain *Feasibility* to assess the reduction of negative economic impacts. For the Influence criterion, we considered the changes to the steady state values of KV or V that belong to target concepts of *'Water Situation (A)'*, *'Water Resources (B)'* and *'Water Demand (C)'* (Fig. 9 A, B and C, respectively). The changes to steady state are normalised and averaged to obtain Influence value. Then the weighted criteria are aggregated and normalized to a percentage value representing "Appropriateness".

We demonstrate the above investigation process at the sub levels of *'Water Projects (F)'* concept (Fig. 10) [28]. This concept consists of 4 key variables and 13 original variables. First, the investigation is made on the level of key variables to rank them according to their "*Appropriateness*". Table 5 shows the values for *Appropriateness* and the corresponding 3 sub criteria for the 4 key variables of the *'Water Projects (F)'* concept.

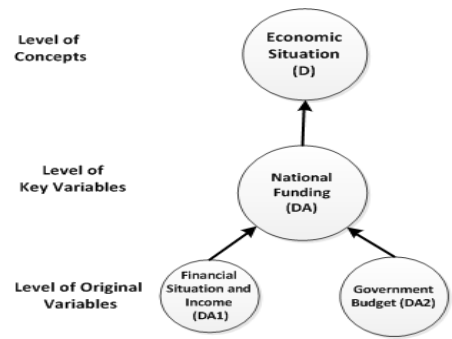

**Fig. 8** Configuration of sub levels of *'Economic Situation (D)'* concept

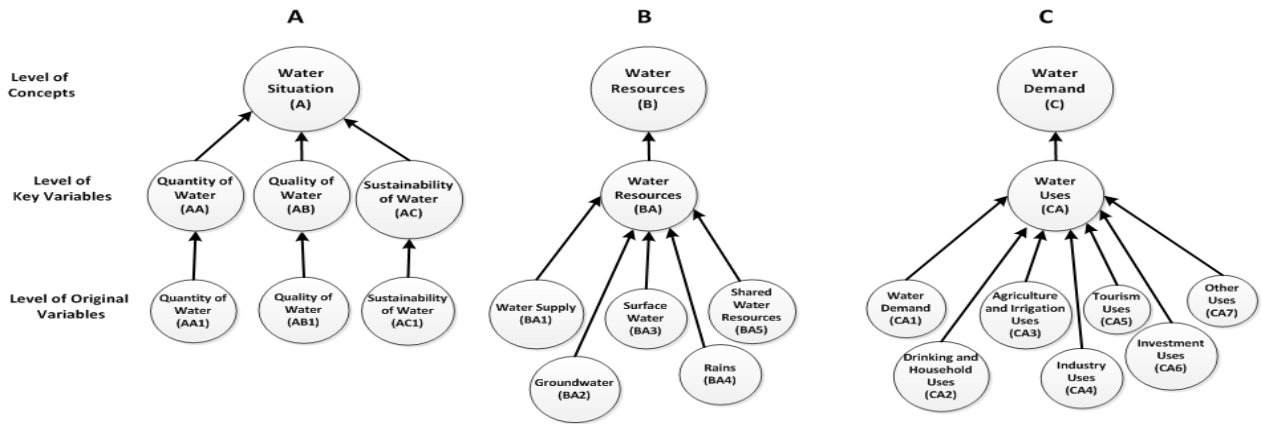

**Fig. 9** Configurations of sub levels of A) *'Water Situation (A)'*, B) *'Water Resources (B)'* and C) *'Water Demand (C)'* concepts.

According to "*Appropriateness*" values in Table 5, *'Water and Rain Harvesting Projects'* (FA) *key variable* ranked the highest. It has high *Importance* and *Influence* values as well as a reasonably small negative economic *Feasibility* value. Although *'Strategic Water Projects'* (FB) key variable representing the two mega water conveyance projects has the highest values for *Importance* and *Influence*, it was ranked second due to its comparatively large negative *Feasibility* value. It is comparatively much more detrimental to the *'Economic Situation (D)'* than other key variables in this concept. The third rank was for *'Water Storage Methods'* (FD) key variable whose *Importance* and *Influence* values are reasonable and its negative *Feasibility* value is very small. *'Develop Water Resources'* (FC) key variable occupied the last rank. Not only its negative *Feasibility* value is slightly large, but also its *Importance* and *Influence* values are small. Accordingly, *'Water and Rain harvesting'* and *'strategic water projects'* are suitable choices for solving the water problem. The *water storage methods* could also help. But *developing water resources* is not cost effective and hence it is inappropriate and ignored from the investigations at the level of original variables.

Next, for each acceptable key variable, the "*Appropriateness*" of its original variables are calculated. As *'Water Storage Methods' key variable* has only one original variable (*'Ground and Surface Reservoirs and Wells'*) (Fig.10), there is no need to calculate its "*Appropriateness*". Table 6 (upper part) presents "*Appropriateness*" of original variables belonging to *'Water and Rain Harvesting Projects'* and *'Strategic Water Projects' key variables* from the scenario simulations at the variable level. As for the analysis of KV, the highest "*Appropriateness*" value is given to *'Wells, Reservoirs & Pits'* (FA3) *original variable*. This is due to its large *Importance* and *Influence* values, as well small negative *Feasibility* value. This variable expresses stakeholders' suggestion to build proper wells, reservoirs and pits to store rainwater. The *'Water Harvesting Projects'* (FA1) occupied the second rank. It means constructing any projects, without specification, to store water. Its *Importance* and *Influence* values are very large, but its negative *Feasibility* value is also very large. *'Proper Dams'* (FA2) occupied the 3rd position. Although the negative *Feasibility* value of *'Water Network for Collecting Rains'* (FA4) is very low, its "*Appropriateness*" is also low due to its lower *Importance* and *Influence* values.

Concerning the original variables of *'Strategic Water Projects'* key variable (Table 6 lower part), the "*Appropriateness*" values of the two mega water conveyance projects (*'Disi'* and *'Red Sea-Dead Sea'*) are close indicating that both are effective. *Disi* is a mega project involving groundwater extraction and distribution over long distances at an estimated cost of US$ 600 million and *'Red Sea-Dead Sea'* is a US$ 10 billion mega project involving extraction of sea water from red sea and discharging it into dead sea during which the water is to be used for power generation and desalination purposes.

**Table 5** "*Appropriateness*" values of key variables belonging to *'Water Projects'* concept (Refer to Fig 8 for the names of KV)

| KV | CW | No of Times Mentioned | Importance | Influence on Econ. Situation | Feasibility | Influence on Water Situation | Influence on Water Resources | Influence on Water Demand | Influence | Appropriateness |
|---|---|---|---|---|---|---|---|---|---|---|
| FA | 28% | 36% | 32% | 17% | 17% | 24% | 35% | 24% | 28% | 35.7% |
| FB | 42.5% | 37.5% | 40% | 60% | -60% | 56% | 36% | 38% | 43% | 33.2% |
| FD | 13.5% | 14.5% | 14% | 1% | -1% | 6% | 4% | 34.5% | 15% | 21.1% |
| FC | 16% | 12% | 14% | 22% | -22% | 14% | 25% | 3.5% | 14% | 10.1% |
| Total | | | 100% | | 100% | | | | 100% | 100% |

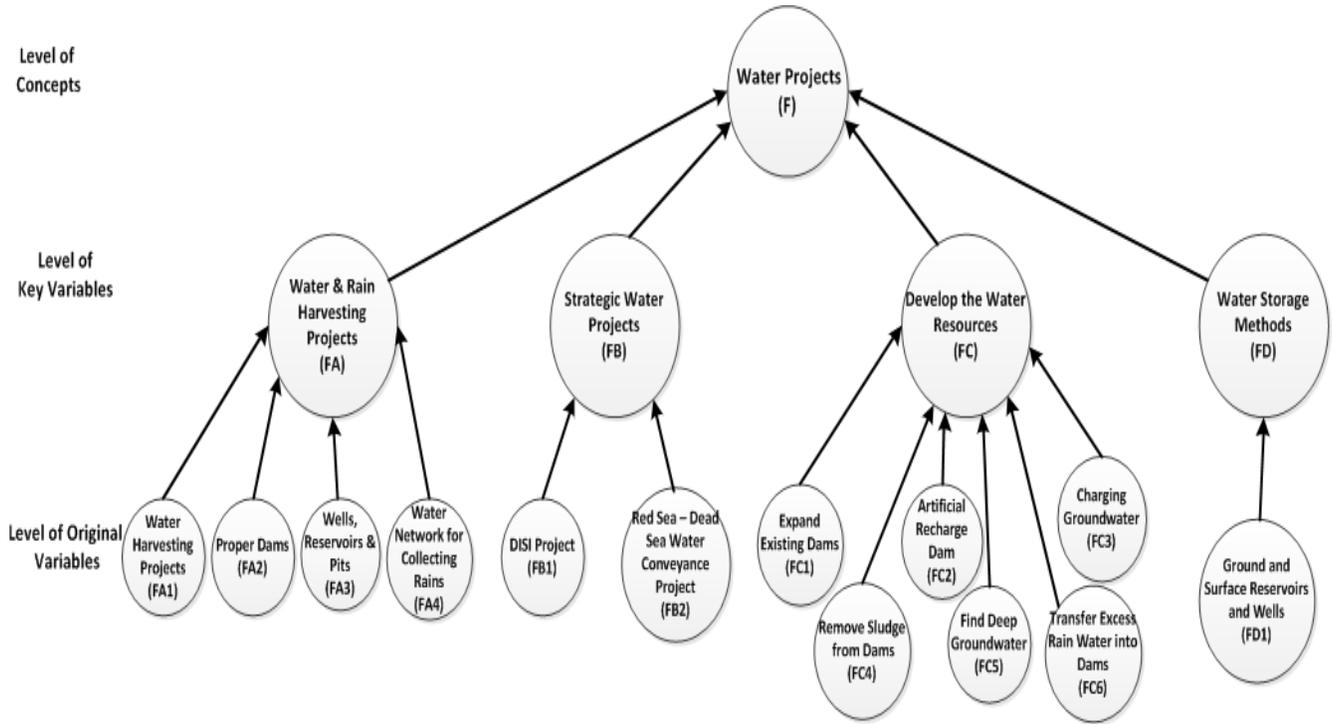

**Fig. 10** Configuration of the sub levels of *'Water Projects'* concept

The negative *Feasibility* value of '*Disi Project'* (FB2) (south to north groundwater conveyance) is the larger of the two because although the cost of operation of *'Red Sea-to-Dead Sea (Conveyance) Project'* (FB1) (for electricity generation and desalination) is much higher, its long term revenue is also greater. However, stakeholders believed that *'Disi Project'* is more realistic than the other and hence the former was mentioned more than the latter in the stakeholder FCMs giving it a higher *Importance* value. Thus construction of water projects plays an important role in solving the water problem. Although some projects such as the two mega water conveyance projects have large negative economic impacts, they show strong positive influence on water situation and water resources while reducing the demand. Other water projects, such as *water and rain harvesting* projects, could also improve the situation to a lesser extent with smaller negative economic impact. From these, drilling suitable wells and pits and constructing reservoirs to store rainwater are very useful. *Constructing proper dams* also could help. Finally, water storage methods, such as *small wells and tanks* in homes and farms to store water from municipal water networks and rainwater could help a little, particularly in reducing water demand.

**Table 6** "*Appropriateness*" values of original variables belonging to *'Water and Rain Harvesting Projects'* and *'Strategic Water Projects'* key variables

| Original Variable | CW (%) | No of Times Mentioned (%) | *Importance* (%) | Influence on Economic Situation (%) | *Feasibility* (%) | Influence on Water Situation (%) | Influence on Water Resources (%) | Influence on Water Demand (%) | *Influence* (%) | *Appropriateness* (%) |
|---|---|---|---|---|---|---|---|---|---|---|
| | | | | Water and Rain Harvesting Projects | | | | | | |
| FA3 | 20 | 36 | 28 | 2 | -2 | 23 | 11.5 | -69.5 | 35 | 48 |
| FA1 | 41 | 40 | 40.5 | 55 | -55 | 39 | 51.5 | 7 | 32 | 25 |
| FA2 | 28 | 21 | 24.5 | 4% | -42 | 28 | 32 | 23 | 28 | 19 |
| FA4 | 11 | 3 | 7 | 1 | -1 | 10 | 5 | 0.5 | 5 | 8 |
| Total | | | 100 | | 100 | | | | 100 | 100 |
| | | | | Strategic Water Projects | | | | | | |
| FB2 | 50 | 45 | 47.5 | 49 | -49 | 54.5 | 54.5 | -53.5 | 54 | 53 |
| FB1 | 50 | 55 | 52.5 | 51 | -51 | 45.5 | 45.5 | 46.5 | 46 | 47 |
| Total | | | 100 | | 100 | | | | 100 | 100 |

FA3: Wells, Reservoirs & Pits for Collecting Rain, FA1: Water Harvesting Projects, FA2: Proper Dams, FA4: Water Network for Collecting Rain, FB1: Disi Project, FB2: Red Sea-Dead Sea Water Conveyance Project

We repeated the above investigation process into the sub levels of *'Integrated Management and Laws (G)'*, *'Technology (I)'*

and *'Community Participation (K)'* concepts to determine the "*Appropriateness*" values of *key variables* and *original variables* belonging to them and selected the most effective original variables that could mitigate water scarcity. These were geared towards maintaining and enhancing existing water resources and managing water demand. (Refer to Figs. S3, S4 and S5 for the Key variables and Variables belonging to these concepts). Regarding *'Integrated Management and Laws'* policy, *'integrated efforts of institutions'* was the most influential original variable and it means that the efforts of all water-related institutions should be integrated to address the problem. Other influential variables belonging to this concept were *'Stable Policies'*, *'Good Management and Institutional Reform'*, *'Skilled Human and Labour Resources'*, *'Laws and By-Laws concerning Water and Water Rights'*, *'Enforcement of Regulations and Laws'*, *'Water Pricing Policy'* and *'Securing Water Rights from Neighbouring Countries'*. Regarding the *'Technology (I)'* policy concept, '*Wastewater Treatment Technology'* was the most influential variable. *'Desalination Technology'*, *'Modern Techniques and Devices'*, *'Nuclear Energy Technology'* and *'Scientific Research'* variables were also somewhat effective. Finally, for the *'Community Participation (K)'* policy, *'Strategic Shift in Agriculture'*, *'Improve Irrigation Systems'*, *'Private Sector Participation'* and *'Regional and International Cooperation'* were the important variables that could help mitigate water scarcity [28].

## 7. Summary, conclusions and recommendations

Most socioecological problems require human participation along with practical system models to generate practical solutions. This research contributed to this goal by demonstrating the efficacy of our advanced semi-quantitative FCM model proposed in [27] with further enhancements for addressing complex participatory real life problems. The efficacy of the model was demonstrated through its application for generating solutions to "*Mitigating Water Scarcity in Jordan*". The study involved development, condensation and aggregation of FCMs of five stakeholder groups at 2 levels of condensation within a fuzzy 2-tuple framework. This was followed by five targeted policy simulations at the 3 levels of FCMs for selecting the most feasible solutions at the original variables level based on a newly proposed fuzzy *Appropriateness* criterion.

The study recommends specific water projects which, although costly, can be considered as new key water resources that can significantly improve or completely solve the water crisis in Jordan. The study also makes a number of other recommendations towards maintaining and enhancing existing water resources and managing water demand. Specifically, it has proposed a number of recommendations as mentioned above along the different themes represented by the four selected important policy concepts of *'Water Projects (F)'*, *'Integrated Management and Laws (G)'*, *'Technology (I)'* and *'Community Participation (K)'*. Finally, this study can be considered as the first attempt at investigating and generating recommendations for implementing the very complex *Jordan Water Strategy* through a comprehensive and systematic study based on proper representation of the varied stakeholder perceptions and sound processing of these perceptions through the rigorous steps of an advanced FCM model that contributed to enhancing the confidence in the generated recommendations. In the implementation of the model, all formulae for condensation, aggregation and other processes and all computations were thoroughly checked to ensure their correctness and model findings were thoroughly assessed and corroborated with stakeholder perceptions and documented literature. All results presented here have received this rigorous scrutiny. These steps are important as they are aspects of validation of FCMs since direct validation of FCMs in a classical sense is not possible. In complex socioecological systems models, such rigour, correctness, reflection, representativeness and validity of the processes involved in the model are the key to ensuring the validity of the outcomes. Therefore, *Water Authorities* in Jordan may consider these recommendations seriously in implementing their *Water Strategy*.

### Software/Data/Availability

All calculations in this study were carried out using MATLAB software. Each FCM, developed by stakeholders, was encoded into a matrix saved in a separate Excel file. This resulted in 123 Excel files in total as detailed in Table S7 and these can be made available on request.

### Supplementary materials:

Available online.

**Acknowledgements:** Authors gratefully acknowledge funding support from Lincoln University, New Zealand.

# Supplementary materials

# Advanced Fuzzy Cognitive Map Framework for Socioecological Systems: Policy development for addressing complex real-life Challenges

## Mamoon Obiedat and Sandhya Samarasinghe

**Table S1** Water supply in Jordan

| | | Source | MCM/Year |
|---|---|---|---|
| Conventional | Renewable water | Surface water (50%) | 505 |
| | | Groundwater (27%) | 275 |
| | Non-renewable | Fossil groundwater (12%) | 125 |
| Non-conventional | Treated water | Wastewater (10%) | 100 |
| | | Desalination (1%) | 10 |
| | | Total | 1015 |

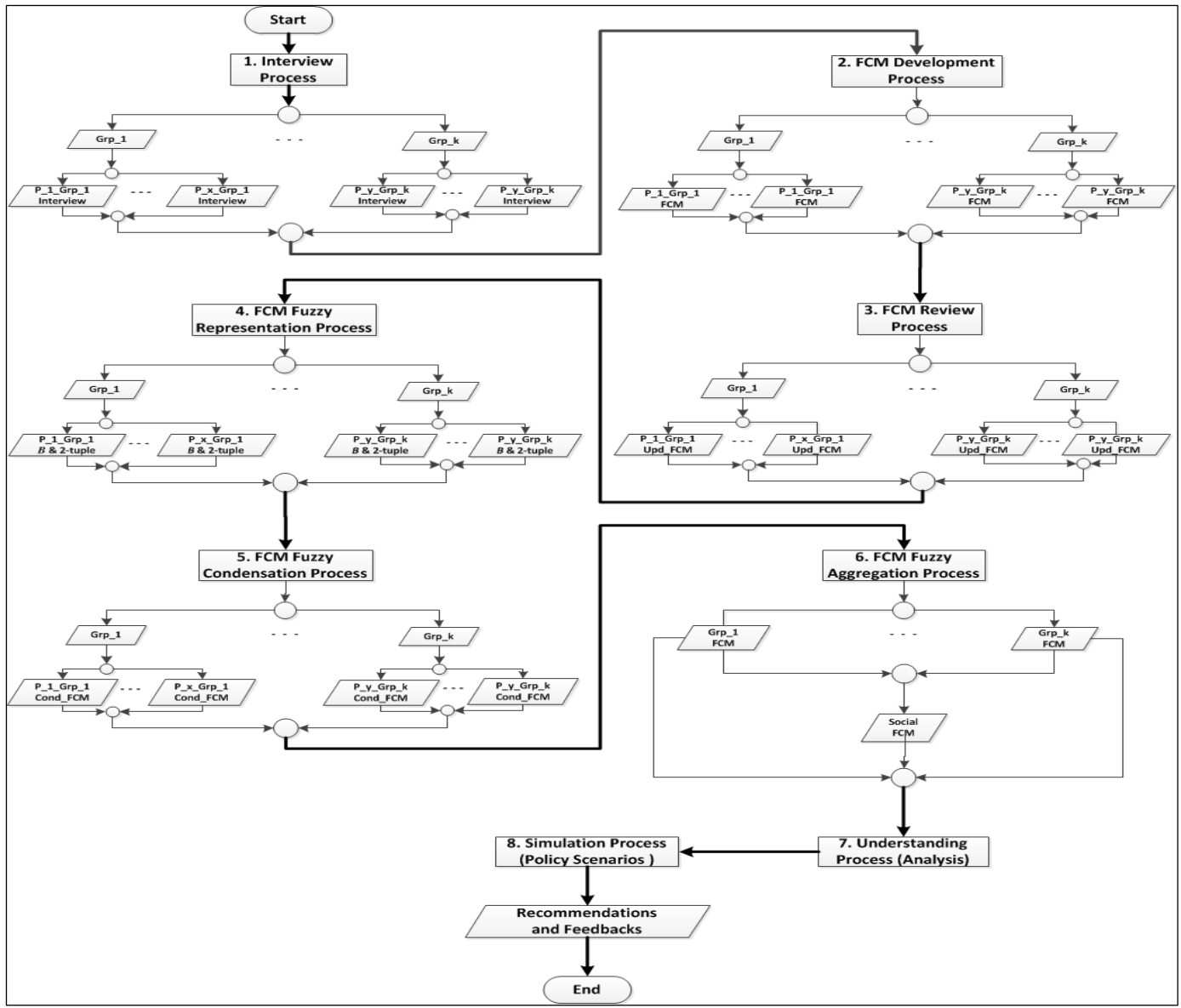

**Fig. S1** The eight step process of the semi-quantitative FCM model [27]

**Table S2** Questionnaire for the interviews

| | |
|---|---|
| A. General questions | 1. What are your age, gender and occupation? |
| | 2. What do you know about general water situation in Jordan and what is behind it? |
| | 3. What can you tell me about water scarcity in Jordan? |
| | 4. Have you any experienced water interruption in your house, if so, how many times, approximately per month and what do you do to manage this problem? |
| B. Questions identifying main factors and key considerations | 1. What are the important conventional water resources (keywords: rainwater, groundwater, surface water, and water shared with neighboring countries) and non-conventional water resources (keywords: water produced from the desalination and the treatment of wastewater and grey water) in Jordan? |
| | 2. What are the main water uses and allocations for these uses in Jordan (keywords: domestic water, irrigation water and other uses)? |
| | 3. Tell me about negative impacts that rise from water scarcity problem in Jordan (keywords: social, investment, economic, political impacts, etc.)? |
| | 4. What are the main factors that affect water scarcity and raise the demand for water in Jordan (keywords: weakness of integrated management of water resources, climate change, high population growth and influxes of refugees, financial constraints, economic growth, lack of water resources, overuse and overexploitation of water resources, development of agriculture, investment and industry, , weakness of laws and by-laws concerning water and water rights, shortage skilled human resources, pollution, wastage of water, and political difficulties)? |
| | 5. What are the most important actions that citizens can take to reduce water scarcity and improve water situation in Jordan (water supply) in Jordan (keywords: surface roof water harvesting, rationalize water consumption by using modern devices, improve individual behaviour such as placing public-interest on private-interest, pay water tariffs without delay, promote confidence between the citizen and the official, and adaptation to the severe shortage of water)? |
| | 6. What are the most important actions that government can take to reduce water scarcity and improve water situation in Jordan (keywords: suitable water system, good management, effective institutional reform, improve and activate water laws and regulations, periodic maintenance and immediately repair the leak and loss of water, control groundwater wells and prevent their overuse, updating water tariffs based on quality and use, promote confidence between the citizen and the official, improve water distribution, seek for new water resources, securing Jordan rights of the water shares with neighbouring countries, increase water harvesting methods, and awareness and educational programs for citizen and farmers)? |
| | 7. What are the most important actions that farmers can take to reduce water scarcity and improve water situation in Jordan (keywords: improve irrigation systems by using more effective irrigation methods, strategic shift in agriculture such as selection of crops varieties with lower water consumption, reduce water losses by implementing water demand management practices and immediately repair the leak and loss of water in their farms, and make pits and wells to collect and store water)? |
| | 8. What is the role of modern techniques such as wastewater and grey water treatment, desalination, the use of modern techniques in irrigation and rationalize consumption, research, information technology and databases to improve and manage the water situation and resources in Jordan? And what are the impacts of these technologies on the financial situation? |
| | 9. Do you think that the participation of the private sector in the water sector would improve the water situation and reduce the scarcity problem and achieve a balance between water supply and demand? And is there should be a hand of government to oversee the private sector in order to secure the rights of people from their basic needs of water? |
| | 10. What are the impacts of the development and urbanization on water situation and financial situation in Jordan? |
| | 11. What do you think about the strategic projects, such as Disi and Red Sea – Dead Sea Water Conveyance to improve the water situation in Jordan? Are they good alternatives to reduce the gap between water demand and supply compared to the large cost of their completion? |
| | 12. Do you think that improving water resources through what previously stated is sufficient to meet all the needs of water and achieve a balance between water demand and supply or there is a need for new sources? If so, what do you suggest for this and what is its impact on the financial situation? |
| | 13. What are the things that you propose for the sustainability of water resources in Jordan (Keywords: traditional knowledge, technology, good management, proper water system, deterrent laws, and reduce wastage of water)? |
| | 14. Is there anything else you would like to say or add to the interview? |

- These questions may be accompanied by sub-questions for further clarification of anything unclear.
- The Interview will finish when the interviewee has nothing to add.

**Table S3** The adjacency (original weights) matrix of FCM in Fig 4 showing the imprecise connection weights and their signs

| Node | $c_1$ | $c_2$ | $c_3$ | $c_4$ | $c_5$ | $c_6$ | $c_7$ | $c_8$ | $c_9$ | $c_{10}$ | $c_{11}$ | $c_{12}$ | $c_{13}$ |
|---|---|---|---|---|---|---|---|---|---|---|---|---|---|
| $c_1$ | 0 | 0 | 0 | 0.37 | 0 | 0 | 0 | 0.68 | 0 | 0 | 0 | -0.71 | -0.15 |
| $c_2$ | 0.76 | 0 | 0 | 0 | 0 | 0 | 0 | 0.27 | 0 | 0 | 0 | -0.31 | -0.26 |
| $c_3$ | -0.71 | -0.13 | 0 | 0 | 0 | 0 | 0 | -0.36 | 0 | 0 | 0 | 0.26 | 0 |
| $c_4$ | 0 | 0.04 | 0 | 0 | 0 | 0.7 | 0.61 | 0.11 | 0.62 | 0 | 0 | 0 | 0 |
| $c_5$ | -0.63 | 0 | 0 | -0.34 | 0 | 0 | 0 | 0 | 0 | 0 | 0 | 0 | 0 |
| $c_6$ | 0.58 | 0.3 | -0.33 | -0.48 | 0 | 0 | 0 | 0 | 0 | 0 | 0 | 0 | 0 |
| $c_7$ | 0.48 | 0.29 | -0.17 | 0 | -0.62 | 0.62 | 0 | 0 | 0.55 | 0.34 | 0.37 | -0.18 | -0.58 |
| $c_8$ | -0.1 | 0 | 0.55 | 0.48 | 0 | 0 | 0 | 0 | 0 | 0 | 0 | 0 | 0 |
| $c_9$ | 0.48 | 0.27 | -0.27 | -0.47 | -0.22 | 0.25 | 0.5 | 0.14 | 0 | 0 | 0.35 | 0 | 0 |
| $c_{10}$ | 0 | 0 | 0 | 0 | -0.09 | 0.32 | 0.2 | 0 | 0 | 0 | 0.41 | 0 | 0 |
| $c_{11}$ | 0 | 0 | -0.47 | 0 | -0.55 | 0.26 | 0.58 | 0 | 0.43 | 0.24 | 0 | 0 | -0.46 |
| $c_{12}$ | 0 | -0.21 | 0 | 0 | 0 | 0 | 0 | -0.51 | 0 | 0 | 0 | 0 | 0.26 |
| $c_{13}$ | -0.17 | -0.63 | 0.63 | -0.34 | 0.31 | 0 | 0 | 0 | 0 | 0 | -0.17 | 0.42 | 0 |

**Table S4** The fuzzy $B$ values representing the adjacency matrix of FCM connection weights shown in Table S2

| Node | $c_1$ | $c_2$ | $c_3$ | $c_4$ | $c_5$ | $c_6$ | $c_7$ | $c_8$ | $c_9$ | $c_{10}$ | $c_{11}$ | $c_{12}$ | $c_{13}$ |
|---|---|---|---|---|---|---|---|---|---|---|---|---|---|
| $c_1$ | 0 | 0 | 0 | 2.21 | 0 | 0 | 0 | 4.09 | 0 | 0 | 0 | -4.27 | -0.87 |
| $c_2$ | 4.54 | 0 | 0 | 0 | 0 | 0 | 0 | 1.60 | 0 | 0 | 0 | -1.89 | -1.59 |
| $c_3$ | -4.28 | -0.77 | 0 | 0 | 0 | 0 | 0 | -2.18 | 0 | 0 | 0 | 1.57 | 0 |
| $c_4$ | 0 | 0.24 | 0 | 0 | 0 | 4.21 | 3.65 | 0.64 | 3.69 | 0 | 0 | 0 | 0 |
| $c_5$ | -3.76 | 0 | 0 | -2.07 | 0 | 0 | 0 | 0 | 0 | 0 | 0 | 0 | 0 |
| $c_6$ | 3.44 | 1.78 | -2.01 | -2.89 | 0 | 0 | 0 | 0 | 0 | 0 | 0 | 0 | 0 |
| $c_7$ | 2.85 | 1.78 | -0.98 | 0 | -3.69 | 3.72 | 0 | 0 | 3.27 | 2.07 | 2.22 | -1.04 | -3.47 |
| $c_8$ | -0.61 | 0 | 3.27 | 2.88 | 0 | 0 | 0 | 0 | 0 | 0 | 0 | 0 | 0 |
| $c_9$ | 2.89 | 1.63 | -1.63 | -2.80 | -1.32 | 1.47 | 3.02 | 0.80 | 0 | 0 | 2.10 | 0 | 0 |
| $c_{10}$ | 0 | 0 | 0 | 0 | -0.56 | 1.95 | 1.16 | 0 | 0 | 0 | 2.48 | 0 | 0 |
| $c_{11}$ | 0 | 0 | -2.84 | 0 | -3.30 | 1.49 | 3.45 | 0 | 2.60 | 1.43 | 0 | 0 | -2.76 |
| $c_{12}$ | 0 | -1.22 | 0 | 0 | 0 | 0 | 0 | -3.03 | 0 | 0 | 0 | 0 | 1.55 |
| $c_{13}$ | -1.01 | -3.78 | 3.76 | -2.06 | 1.90 | 0 | 0 | 0 | 0 | 0 | -0.98 | 2.51 | 0 |

**Table S5** Values of Degree, Closeness, Betweenness and Consensus centrality and $CW$ of nodes in FCM in Fig 1

| Node | Degree Centrality | Closeness Centrality | Betweenness Centrality | CCM | Credibility weight ($cw_i$) |
|---|---|---|---|---|---|
| $c_1$ | 5.66 | 4.12 | 1.27 | 3.69 | 0.113 |
| $c_2$ | 2.52 | 5.45 | 2.81 | 3.59 | 0.11 |
| $c_3$ | 3.06 | 0.71 | 0.38 | 1.38 | 0.042 |
| $c_4$ | 3.95 | 3.11 | 2.56 | 3.21 | 0.098 |
| $c_5$ | 1.56 | 0.75 | 0.5 | 0.94 | 0.029 |
| $c_6$ | 2.99 | 0 | 0 | 1 | 0.031 |
| $c_7$ | 6 | 1.74 | 0.5 | 2.75 | 0.084 |
| $c_8$ | 2.14 | 3.6 | 1.13 | 2.29 | 0.07 |
| $c_9$ | 3.93 | 3.17 | 0 | 2.37 | 0.072 |
| $c_{10}$ | 0 | 2.28 | 1.27 | 1.18 | 0.036 |
| $c_{11}$ | 3.58 | 3.78 | 3.56 | 3.64 | 0.112 |
| $c_{12}$ | 1.68 | 2.37 | 0.25 | 1.43 | 0.044 |
| $c_{13}$ | 3.71 | 6 | 6 | 5.24 | 0.159 |

**Table S6** Some indices of the group FCMs at the level of the concepts

| Group | CCM | No of Concepts | No of Connections | Density | 1st Important Concept | 2nd Important Concept | 3rd Important Concept | 4th Important Concept | 5th Important Concept |
|---|---|---|---|---|---|---|---|---|---|
| Private Sector | 0.601 | 13 | 89 | 0.570 | M | G | D | I | F |
| Local People | 0.566 | 13 | 92 | 0.590 | G | M | A | H | D |
| Experts | 0.612 | 13 | 102 | 0.654 | M | H | I | A | C |
| Managers | 0.601 | 13 | 100 | 0.641 | H | M | G | A | D |
| Farmers | 0.588 | 13 | 100 | 0.641 | H | M | K | G | B |

Average ± SD; (min; max) of CCM of the 5 group FCMs  0.593±0.018;  (min=0.566; max=0.612)

A: Water Situation, B: Water Resources, C: Water Demand, D: Economic Situation, E: Wastage of Water, F: Water Projects, G: Integrated Management and Laws, H: Development & Urbanization, I: Technology, J: Acquired and Conventional Knowledge, K: Community Participation, L: Consequences of Water Scarcity, M: Causes of Water

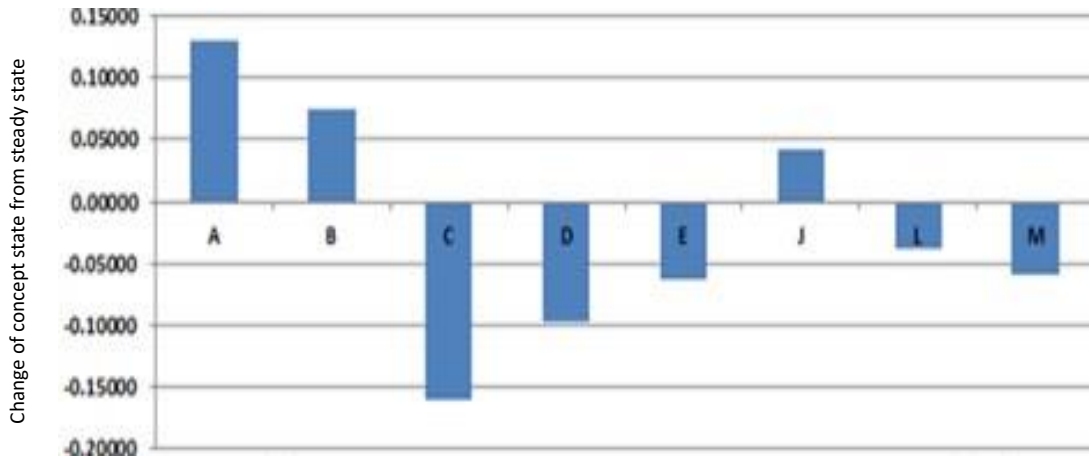

**Fig. S2.A** Change in steady state of nodes from integration of five policy scenarios into one scenario.

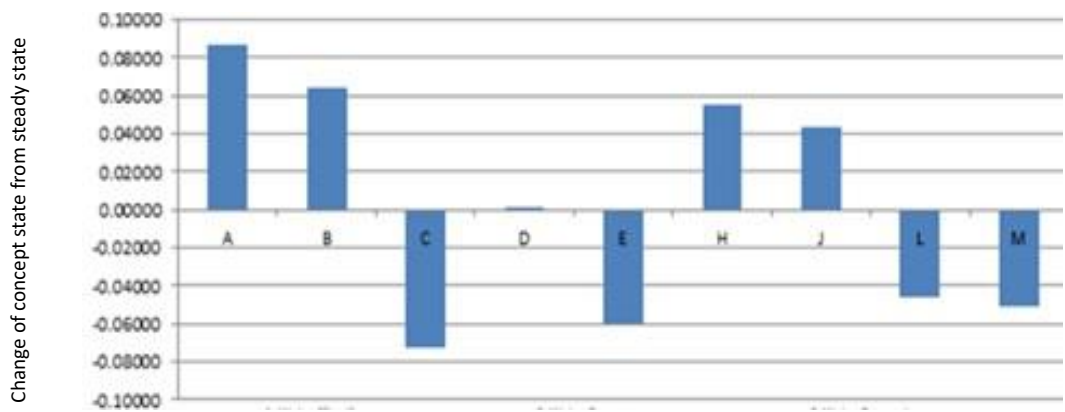

**Fig. S2.B** Change in steady state of nodes from integration of all but *'Development and Urbanization'* into one scenario

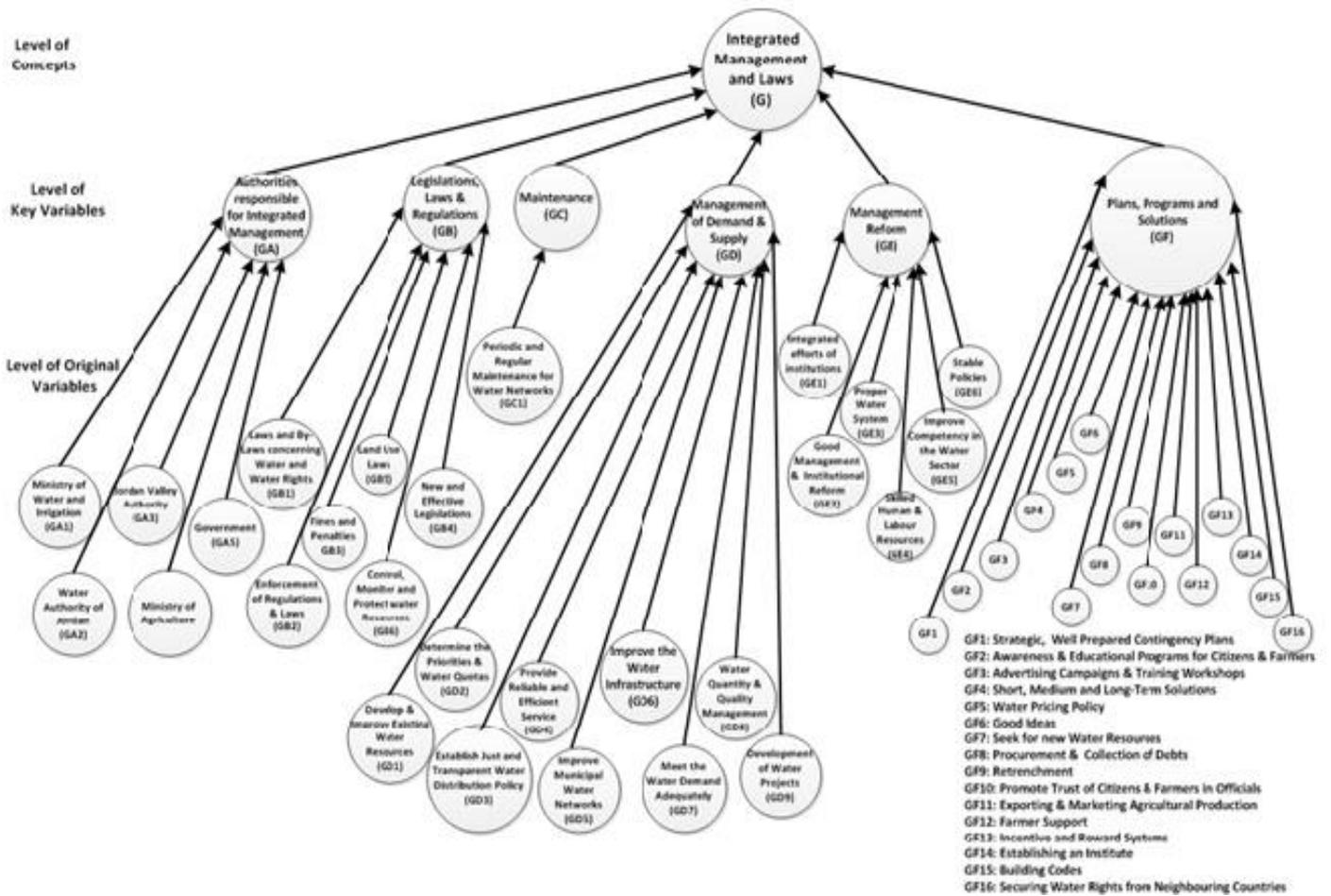

**Fig. S3** Configuration of the sub levels of *'Integrated Management and Laws'* concept

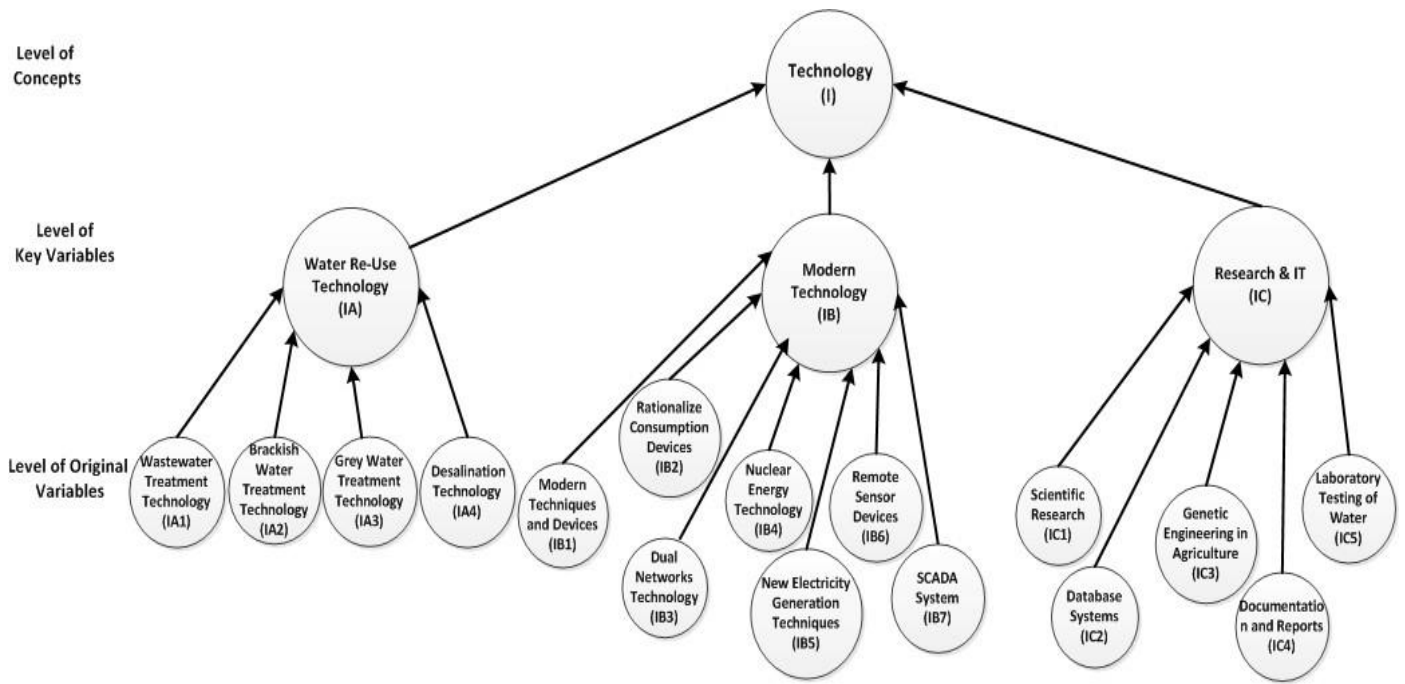

**Fig. S4** Configuration of the sub levels of *'Technology'* concept

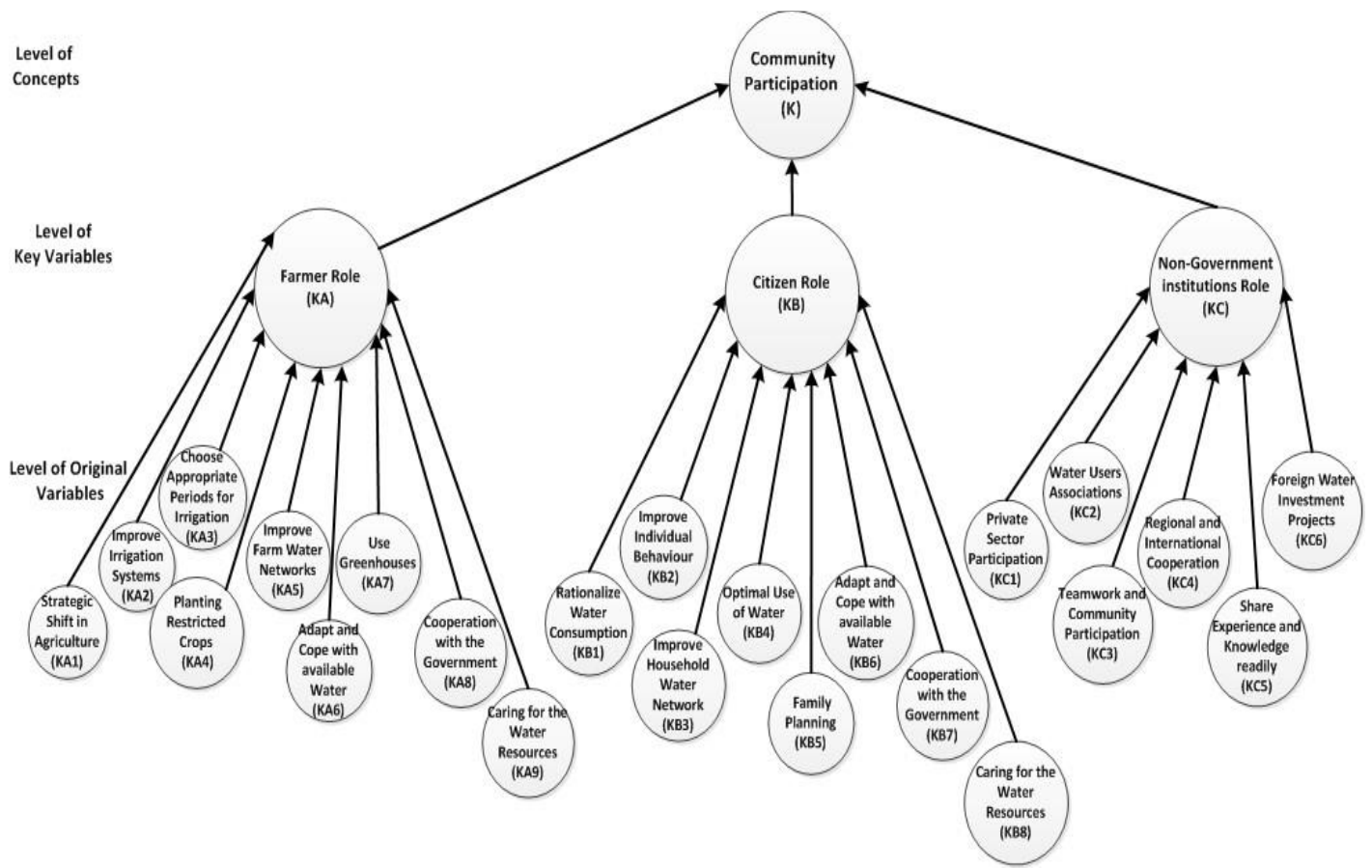

**Fig. S5** Configuration of the sub levels of *'Community Participation'* concept

**Table S7** Description of datasets used in the study

| FCM level | # of Excel files | Description |
| --- | --- | --- |
| Original variables | 35 | Resulted from stakeholders |
| | 5 | Resulted from the aggregation process for each group at the level of original variables |
| | 1 | Resulted from the aggregation process for all FCMs at the level of original variables |
| Key variables | 35 | Resulted from the condensation process at the 1st level of condensation (level of key variables) |
| | 5 | Resulted from the aggregation process for each group at the level of key variables |
| | 1 | Resulted from the aggregation process for all FCMs at the level of key variables |
| Concepts | 35 | Resulted from the condensation process at the 2nd level of condensation (level of concepts) |
| | 5 | Resulted from the aggregation process for each group at the level of concepts |
| | 1 | Resulted from the aggregation process for all FCMs at the level of concepts |
| Total | 123 | |

**Table S8** The two levels of condensation, level one includes the Key Variables, and level two includes the Key Concepts

| Key Considerations | | Key Variables | | | Original Variables | |
|---|---|---|---|---|---|---|
| Code | Name | Code | | Name | Code | Name |
| A | 1. Water Situation | AA | 1. | Quantity of Water | AA1 | 1. Quantity of Water |
| | | AB | 2. | Quality of Water | AB1 | 2. Quality of Water |
| | | AC | 3. | Sustainability of Water | AC1 | 3. Sustainability of Water |
| B | 2. Water Resources | BA | 4. | Water Resources | BA1 | 4. Water Supply |
| | | | | | BA2 | 5. Ground Water |
| | | | | | BA3 | 6. Surface Water |
| | | | | | BA4 | 7. Rains |
| | | | | | BA5 | 8. Shared Water Resources |
| C | 3. Water Demand | CA | 5. | Water Uses (Demand) | CA1 | 9. Water Demand |
| | | | | | CA2 | 10. Drinking & Household Uses |
| | | | | | CA3 | 11. Agriculture and Irrigation Uses |
| | | | | | CA4 | 12. Industry Uses |
| | | | | | CA5 | 13. Tourism Uses |
| | | | | | CA6 | 14. Investment Uses |
| | | | | | CA7 | 15. Other Uses |
| D | 4. Economic Situation | DA | 6. | National Funding | DA1 | 16. Financial Situation and Income |
| | | | | | DA2 | 17. Government Budget |
| | | DB | 7. | External Funding | DB1 | 18. Donor Countries |
| | | | | | DB2 | 19. Foreign Loans |
| | | DC | 8. | Farmer Funding | DC1 | 20. Farmers' Financial Situation |
| | | | | | DC2 | 21. Farmers' Agricultural Loans |
| E | 5. Wastage of Water | EA | 9. | Loss of Water Networks | EA1 | 22. Technical Losses & Leaking of the Water Networks |
| | | | | | EA2 | 23. Administrative Losses |
| | | EB | 10. | Wastage of irrigation water | EB1 | 24. Water Losses During Irrigation |
| | | | | | EB2 | 25. Evaporation that could be Controlled |
| | | EC | 11. | Wastage of Rainfall | EC1 | 26. Wastage of Rainfall |
| F | 6. Water Projects | FA | 12. | Water & Rain Harvesting Projects | FA1 | 27. Water Harvesting Projects |
| | | | | | FA2 | 28. Proper Dams |
| | | | | | FA3 | 29. Wells, Reservoirs & Pits for Collecting Rain |
| | | | | | FA4 | 30. Water Network for Collecting Rains |
| | | FB | 13. | Strategic Water Projects | FB1 | 31. Disi Project |
| | | | | | FB2 | 32. Red Sea – Dead Sea Water Conveyance Project |
| | | FC | 14. | Develop the Water Resources | FC1 | 33. Expand Existing Dams |
| | | | | | FC2 | 34. Artificial Recharge Dam |
| | | | | | FC3 | 35. Charging Groundwater |
| | | | | | FC4 | 36. Remove Sludge from Dams |
| | | | | | FC5 | 37. Find Deep Groundwater |
| | | | | | FC6 | 38. Transfer Excess Rain Water into Dams |
| | | FD | 15. | Water Storage Methods | FD1 | 39. Ground and Surface Reservoirs and Wells |
| G | 7. Integrated Management and Laws | GA | 16. | Authorities responsible for Integrated Management | GA1 | 40. Ministry of Water and Irrigation |
| | | | | | GA2 | 41. Water Authority of Jordan |
| | | | | | GA3 | 42. Jordan Valley Authority |
| | | | | | GA4 | 43. Ministry of Agriculture |
| | | | | | GA5 | 44. Government |
| | | GB | 17. | Legislations, Laws & Regulations | GB1 | 45. Laws and By-Laws concerning Water and Water Rights |
| | | | | | GB2 | 46. Enforcement of Regulations & Laws |
| | | | | | GB3 | 47. Fines and Penalties |
| | | | | | GB4 | 48. New and Effective Legislations and Powerful Policies. |
| | | | | | GB5 | 49. Land Use Laws |
| | | | | | GB6 | 50. Control, Monitor and Protect water Resources and Equipment |
| | | GC | 18. | Maintenance | GC1 | 51. Periodic and Regular Maintenance for Water Networks |
| | | GD | 19. | Management of Demand & Supply | GD1 | 52. Develop & Improve Existing Water Resources |
| | | | | | GD2 | 53. Determine the Priorities & Water Quotas |
| | | | | | GD3 | 54. Establish Just and Transparent Water Distribution Policy |
| | | | | | GD4 | 55. Provide Reliable and Efficient Service |
| | | | | | GD5 | 56. Improve Municipal Water Networks |
| | | | | | GD6 | 57. Improve the Water Infrastructure |
| | | | | | GD7 | 58. Meet the Water Demand Adequately |
| | | | | | GD8 | 59. Water Quantity & Quality Management |
| | | | | | GD9 | 60. Development of Water Projects |
| | | | | | GE1 | 61. Integrated efforts of institutions |
| | | | | | GE2 | 62. Good Management & Institutional Reform |
| | | | GE | 20. Management Reform | GE3 | 63. Proper Water System |
| | | | | | GE4 | 64. Skilled Human & Labour Resources |
| | | | | | GE5 | 65. Improve Competency in the Water Sector |
| | | | | | GE6 | 66. Stable Policies |
| | | | | | GF1 | 67. Strategic, Well Prepared Contingency Plans |
| | | | | | GF2 | 68. Awareness & Educational Programs for Citizens & Farmers |
| | | | | | GF3 | 69. Advertising Campaigns & Training Workshops |
| | | | | | GF4 | 70. Short, Medium and Long-Term Solutions |
| | | | | | GF5 | 71. Water Pricing Policy |
| | | | | | GF6 | 72. Good Ideas |
| | | | | | GF7 | 73. Seek for new Water Resources |
| | | | GF | 21. Plans, Programs and Solutions | GF8 | 74. Procurement & Collection of Debts |
| | | | | | GF9 | 75. Retrenchment |
| | | | | | GFZ0 | 76. Promote Trust of Citizens & Farmers in Officials |
| | | | | | GFZ1 | 77. Exporting & Marketing Agricultural Production |
| | | | | | GFZ2 | 78. Farmer Support |
| | | | | | GFZ3 | 79. Incentive and Reward Systems |
| | | | | | GFZ4 | 80. Establishing an Institute |
| | | | | | GFZ5 | 81. Building Codes |
| | | | | | GFZ6 | 82. Securing Water Rights from Neighbouring Countries |
| | | | | | HA1 | 83. General Development |
| | | | | | HA2 | 84. Agricultural Development |
| | | | | | HA3 | 85. Industrial Development |
| | | | | | HA4 | 86. Tourism Development |
| | | | | | HA5 | 87. Investment Development |
| | | | HA | 22. Development | HA6 | 88. Human Resources Development |
| H | 8. | Development and Urbanization | | | HA7 | 89. Economic Growth and Development |
| | | | | | HA8 | 90. Farmland Expansion and Forestation |
| | | | | | HA9 | 91. Level of Dead Sea ( Reviving the Dead Sea) |
| | | | | | HA10 | 92. Improve Food and Health Security |
| | | | | | HA11 | 93. Agricultural Productivity |
| | | | | | HA12 | 94. Desert Farming |
| | | | | | HB1 | 95. General Urbanization |
| | | | HB | 23. Urbanization | HB2 | 96. Life Style Improvement |
| | | | | | HB3 | 97. Expansion of Construction |
| | | | | | HB4 | 98. Greening & Landscaping the Country |
| | | | | | IA1 | 99. Wastewater Treatment Technology |
| | | | IA | 24. Water Re-Use Technology | IA2 | 100. Brackish Water Treatment Technology |
| | | | | | IA3 | 101. Grey Water Treatment Technology |
| | | | | | IA4 | 102. Desalination Technology |
| | | | | | IB1 | 103. Modern Techniques and Devices |
| | | | | | IB2 | 104. Rationalize Consumption Devices |
| I | 9. | Technology | IB | 25. Modern Technology | IB3 | 105. Dual Networks Technology |
| | | | | | IB4 | 106. Nuclear Energy Technology |
| | | | | | IB5 | 107. New Electricity Generation Techniques |
| | | | | | IB6 | 108. Remote Sensor Devices |
| | | | | | IB7 | 109. SCADA System |
| | | | | | IC1 | 110. Scientific Research |
| | | | | | IC2 | 111. Database Systems |
| | | | IC | 26. Research & IT | IC3 | 112. Genetic Engineering in Agriculture |
| | | | | | IC4 | 113. Documentation and Reports |
| | | | | | IC5 | 114. Laboratory Testing of Water |
| J | | | JA | 27. Acquired Knowledge | JA1 | 115. Citizen Awareness |

| | | | | | | | |
|---|---|---|---|---|---|---|---|
| | 10. Acquired and Conventional Knowledge | | | | JA2 | 116. | Farmer Awareness |
| | | | | | JA3 | 117. | Societal Awareness & Realization |
| | | | | | JA4 | 118. | Education |
| | | JB | 28. | Conventional | JB1 | 119. | Traditional Knowledge and Understanding |
| K | 11. Community Participation | KA | 29. | Farmer Role | KA1 | 120. | Strategic Shift in Agriculture |
| | | | | | KA2 | 121. | Improve Irrigation Systems |
| | | | | | KA3 | 122. | Choose Appropriate Periods for Irrigation |
| | | | | | KA4 | 123. | Planting Restricted Crops |
| | | | | | KA5 | 124. | Improve Farm Water Networks |
| | | | | | KA6 | 125. | Adapt and Cope with available Water |
| | | | | | KA7 | 126. | Use Greenhouses |
| | | | | | KA8 | 127. | Cooperation with the Government |
| | | | | | KA9 | 128. | Caring for the Water Resources |
| | | KB | 30. | Citizen Role | KB1 | 129. | Rationalize Water Consumption |
| | | | | | KB2 | 130. | Improve Individual Behaviour |
| | | | | | KB3 | 131. | Improve Household Water Network |
| | | | | | KB4 | 132. | Optimal Use of Water |
| | | | | | KB5 | 133. | Family Planning |
| | | | | | KB6 | 134. | Adapt and Cope with available Water |
| | | | | | KB7 | 135. | Cooperation with the Government |
| | | | | | KB8 | 136. | Caring for the Water Resources |
| | | KC | 31. | Non-Governmental institutions Role | KC1 | 137. | Private Sector Participation |
| | | | | | KC2 | 138. | Water Users Associations |
| | | | | | KC3 | 139. | Teamwork and Community Participation |
| | | | | | KC4 | 140. | Regional and International Cooperation |
| | | | | | KC5 | 141. | Share Experience and Knowledge readily |
| | | | | | KC6 | 142. | Foreign Water Investment Projects |
| L | 12. Consequences of Water Scarcity | LA | 32. | Negative Social Impacts | LA1 | 143. | General Social Impacts (Pressure, Stress, Injustice and Social |
| | | | | | LA2 | 144. | Low level of Health and Hygiene |
| | | | | | LA3 | 145. | Motivation to do Crime |
| | | | | | LA4 | 146. | Survival and Fear for the Future |
| | | LB | 33. | Negative Economic Impacts | LB1 | 147. | General Economic Impacts |
| | | | | | LB2 | 148. | Investment Impacts |
| | | | | | LB3 | 149. | Tourism Impacts |
| | | | | | LB4 | 150. | Restriction on Development |
| | | | | | LB5 | 151. | Decline in Livestock |
| | | LC | 34. | Negative Agricultural Impacts | LC1 | 152. | Restriction on Agriculture and Productivity |
| | | | | | LC2 | 153. | Diminishing Farmlands |
| | | | | | LC3 | 154. | High Cost of Agriculture |
| | | | | | LC4 | 155. | Diseases and Destruction of Crops |
| | | LD | 35. | Negative Policy Impacts | LD1 | 156. | Political Problems with Neighbouring Countries |
| | | | | | LD2 | 157. | Wars for Water |
| | | LE | 36. | Negative Environmental Impacts | LE1 | 158. | Drought and Desertification |
| | | | | | LE2 | 159. | Pollution |
| M | 13. Causes of Water Scarcity | MA | 37. | Overpopulation | MA1 | 160. | Incremental and Accelerated Population Growth |
| | | | | | MA2 | 161. | Natural Population Growth |
| | | | | | MA3 | 162. | Short Migrations |
| | | | | | MA4 | 163. | Refugees Influx |
| | | | | | MA5 | 164. | Incoming Expatriates and Visitors |
| | | | | | MA6 | 165. | Foreign Workers |
| | | MB | 38. | International Policies with Neighbours | MB1 | 166. | Political Difficulties and Crises |
| | | | | | MB2 | 167. | Non-Fulfilment of the Rights to Shared Water, Peace Treaty |
| | | MC | 39. | Negative Environmental Effects | MC1 | 168. | Climate Change |
| | | | | | MC2 | 169. | Increase in Decertified and Dried Lands |
| | | | | | MC3 | 170. | Entrapment and Fluctuating Rainfall |
| | | | | | MC4 | 171. | Geographical Situation |
| | | | | | MC5 | 172. | Scarcity of Renewable Water Resources |
| | | MD | 40. | Negative Behaviours | MD1 | 173. | Overexploitation and Depletion of Groundwater |
| | | | | | MD2 | 174. | Overuse and Excessive Use of water |

|    |     |                      | MD3 | 175. | Water Pollution |
|----|-----|----------------------|-----|------|------|
|    |     |                      | MD4 | 176. | Assault on Water Sources |
|    |     |                      | MD5 | 177. | Corruption and Favouritism |
|    |     |                      | MD6 | 178. | Use of Water Illegally |
|    |     |                      | MD7 | 179. | Use of Highlands Water |
|    |     |                      | MD8 | 180. | Excessive use of Fertilizers in Agriculture |
|    |     |                      | ME1 | 181. | Summer Season Demand |
| ME | 41. | Demand Stimulations  | ME2 | 182. | Exceeding Amman's Quota of Irrigation Water |
|    |     |                      | ME3 | 183. | Pumping and Energy Interruptions |
|    |     |                      | ME4 | 184. | Technical Problems |
| MF | 42. | Faults and High Cost | MF1 | 185. | High Cost of Water |
|    |     |                      | MF2 | 186. | Cost of Operation and Maintenance of Water Infrastructure |